\documentclass[%
 aip,
 amsmath,amssymb,
 reprint,%
]{revtex4-1}

\usepackage{graphicx}
\usepackage{dcolumn}
\usepackage{bm}

\usepackage{booktabs}
\usepackage[utf8]{inputenc}
\usepackage[T1]{fontenc}
\usepackage{mathptmx}
\usepackage{etoolbox}
\usepackage{subcaption}
\usepackage{needspace}
\usepackage{siunitx}
\usepackage[english]{babel}

\makeatletter
\def\@email#1#2{%
 \endgroup
 \patchcmd{\titleblock@produce}
  {\frontmatter@RRAPformat}
  {\frontmatter@RRAPformat{\produce@RRAP{*#1\href{mailto:#2}{#2}}}\frontmatter@RRAPformat}
  {}{}
}%
\makeatother
\begin{document}

\preprint{AIP/123-QED}

\title[Stability and optical quality of "windmill"-formed 8CB liquid crystal films for replenishable plasma mirrors
]{Stability and optical quality of "windmill"-formed 8CB liquid crystal films for replenishable plasma mirrors
}
\author{A. Vazquez}
\affiliation{Lawrence Berkeley National Laboratory, Berkeley, California 94720, USA}
\affiliation{Department of Nuclear Engineering, University of California, Berkeley, California 94720, USA}

\author{A. Jewell}
\affiliation{Lawrence Berkeley National Laboratory, Berkeley, California 94720, USA}

\author{M. Cole}
\affiliation{Lawrence Berkeley National Laboratory, Berkeley, California 94720, USA}

\author{A. Abdi}
\affiliation{Lawrence Berkeley National Laboratory, Berkeley, California 94720, USA}

\author{T. K. Le}
\affiliation{Lawrence Berkeley National Laboratory, Berkeley, California 94720, USA}

\author{A. McIlvenny}
\affiliation{Lawrence Berkeley National Laboratory, Berkeley, California 94720, USA}

\author{N. Czapla}
\affiliation{The Ohio State University, Columbus, Ohio 43210, USA}

\author{L. Russell}
\affiliation{Lawrence Berkeley National Laboratory, Berkeley, California 94720, USA}
\affiliation{Department of Nuclear Engineering, University of California, Berkeley, California 94720, USA}

\author{A. Picksley}
\affiliation{Lawrence Berkeley National Laboratory, Berkeley, California 94720, USA}

\author{A. J. Gonsalves}
\affiliation{Lawrence Berkeley National Laboratory, Berkeley, California 94720, USA}

\author{K. Nakamura}
\affiliation{Lawrence Berkeley National Laboratory, Berkeley, California 94720, USA}

\author{D. W. Schumacher}
\affiliation{The Ohio State University, Columbus, Ohio 43210, USA}

\author{Z. Eisentraut}
\affiliation{Lawrence Berkeley National Laboratory, Berkeley, California 94720, USA}

\author{F. Mazzini}
\affiliation{Lawrence Berkeley National Laboratory, Berkeley, California 94720, USA}

\author{C. B. Schroeder}
\affiliation{Lawrence Berkeley National Laboratory, Berkeley, California 94720, USA}
\affiliation{Department of Nuclear Engineering, University of California, Berkeley, California 94720, USA}

\author{J. van Tilborg}
\affiliation{Lawrence Berkeley National Laboratory, Berkeley, California 94720, USA}

\author{J. Osterhoff}
\affiliation{Lawrence Berkeley National Laboratory, Berkeley, California 94720, USA}

\author{E. Esarey}
\affiliation{Lawrence Berkeley National Laboratory, Berkeley, California 94720, USA}

\author{L. Obst-Huebl}
\email{lobsthuebl@lbl.gov} 
\affiliation{Lawrence Berkeley National Laboratory, Berkeley, California 94720, USA}

\date{\today}
             
\begin{abstract}
Liquid crystal (LC) film plasma mirrors (PMs) based on 4-octyl-4’-cyanobiphenyl (8CB) are an enabling technology for reflecting high-fluence laser pulses. These freestanding LC films can achieve high optical quality and are well-suited for rep-rated applications, as motorized devices continuously replenish films over an aperture following each destructive laser shot. However, a systematic characterization of film quality as a function of seminal operating conditions had not yet been performed for the LC "windmill" version of the device, which aims to match the repetition rate of an existing "spinning disk" (SDI) version and the angular stability of the "linear slider" (LSTI) version. We determined the 8CB film quality using low-power wavefront measurements, and studied the film-to-film wavefront stability and formation reliability. The film-formation reliability of 8CB LC films demonstrated 97\% formation success at 2.7 mm/s film-forming speeds, but decreased to 45\% at 10.8 mm/s. These reliability numbers will inform future designs to reach Hz-level repetition rates and beyond. Depending on area-of-interest within the 10 mm diameter film, the added wavefront root-mean-squared (RMS) variation was as small as 10 nm for a 2 mm diameter region, and $<$50 nm for a 3 mm diameter region. Within the optimal 21-22$^\circ$C operating regime, pointing fluctuations remained at or below 0.52 mrad. With a maximum effective film formation frequency of \textasciitilde0.25 Hz, these results establish windmill-formed 8CB films as promising candidates to pursue next-iteration improvements towards rep-rated plasma-mirror operation.

\end{abstract}

\maketitle

\section{\label{sec:level1}Introduction}
Modern high-power laser facilities\cite{Danson2019} can commonly achieve intensities greater than $10^{21}$~W/cm$^2$, with state-of-the-art systems\cite{Yoon2021} recently demonstrating intensities exceeding $10^{23}$~W/cm$^2$. In many applications, such ultra-intense pulses must be delivered using optical systems that can tolerate extreme, localized fluence and are required to operate reliably at high repetition rates.

To overcome this inherent limitation, plasma mirrors (PMs) have emerged as a promising solution.\cite{Thaury2007, Sokollik2010} A PM typically consists of a solid state substrate, the surface of which rapidly ionizes upon the arrival of the laser pulse, transforming into a highly reflective plasma. When operated at laser intensities $10^{15}$-$10^{17}$~W/cm$^2$, PMs can reflect up to 90\% of the beam energy, and in addition to rerouting a high-intensity laser pulse, can also significantly clean the laser pulse's temporal contrast from pre-pulses and pedestals.\cite{Dromey2004PlasmaMirror, Obst2018, Shaw2016} However, after this interaction, the corresponding region of the substrate is destroyed and must be replaced, motivating replenishable PM concepts for continuous operation.

In order to enable continuous, uninterrupted PM operation at the repetition rate of modern high-power laser systems, tape-based solutions have been widely used (e.g., VHS tapes of approximately 10~\textmu m thickness), which spool to a fresh, undamaged location after each shot.\cite{Gruse2025SelfGuidedPlasmaMirrors,Prencipe2017TargetsHighRepetitionRate,Barber2020, Shaw2016} Previous work at the BELLA Center demonstrated that tape drives provide practical replenishable PM operation, reporting a peak reflectance of 82\% at an intensity of $4 \times 10^{17}$~W/cm$^2$ at the tape surface with wavefront (focal mode) quality conserved.\cite{Shaw2016} In the same study, the PM-induced pointing fluctuations were found to be on the order of $\sim$1~mrad (with a substantial contribution of $\sim$310~\textmu rad attributed to spooling mechanics).\cite{Shaw2016} More recently, the use of a thicker 125~\textmu m Kapton tape tensioned over a metal guide was shown to reduce these tape-induced pointing errors to $\sim$0.5~mrad.\cite{Gruse2025SelfGuidedPlasmaMirrors} Even when tension is optimized to mitigate these macroscopic effects, holding the tape consistently remains a limiting factor. The root-mean-square (RMS) wavefront error of the VHS tape surface has been measured to be as low as 14~nm~\cite{Sokollik2010}, with substantial additional contributions depending on the ability of the spooling mechanics to straighten the tape. Beyond these limitations in surface quality, utilizing a tape drive poses risks to the lifetime of surrounding equipment due to the high amount of debris that is generated on each shot.

An important application of replenishable PMs lies in multistage coupling of laser plasma accelerators (LPA)\cite{steinke2016, steinke2016staging, schroeder2023linear}, which represents a promising route to reach the multi-10 GeV electron energies with high rep-rate, lower per-stage laser energy.\cite{Esarey2009,steinke2016} There, an electron bunch produced by an LPA stage is injected into subsequent, independently laser-driven stages to iteratively increase its energy.\cite{Picksley2022,steinke2016} A key challenge in LPA staging is the need to couple additional, high-power laser pulses into subsequent LPA stages where conventional coupling optics would be destroyed by the high laser fluence. In the ongoing multi-GeV staging experiments at the BELLA Center's Petawatt (BELLA PW) laser facility\cite{Nakamura2017}, a first LPA is driven by a laser pulse, producing a GeV-class electron bunch.\cite{Gonsalves2019Petawatt} After transport by an active plasma lens\cite{vanTilborg2015}, the bunch passes through a PM before entering a second LPA stage. A second laser pulse from an independent beamline\cite{Picksley2022} is to be coupled via reflection from the PM onto the electron beam axis, where it drives the second acceleration stage and further accelerates the electron bunch to multi-GeV energies.\cite{Stackhouse2022, Sokollik2010} In this configuration, the performance of the second stage depends strongly on the optical quality of the PM, its formation reliability, and on the post-PM pointing of the coupled laser pulse. Also, PMs can be used to protect the front of the active plasma lens from damage induced by the otherwise transmitted stage-1 drive laser. 

The cumulative emittance growth introduced to the electron bunch by traveling through the PMs should remain as small as possible. Previous work has shown a significant advantage of liquid crystal (LC) PMs over tape-based PMs for emittance preservation: their thickness can be as low as few tens of nanometers, orders of magnitude thinner than the tens of micrometers typical for tapes\cite{Zingale2021}. Zingale \textit{et al.}\cite{Zingale2021} used a single $\sim$20~nm LC PM and measured a downstream normalized emittance of $\sim$4~$\mu$m for a transmitted 0.84~GeV electron beam, while estimating (via a scattering model and simulations) that the LC PM contribution from electron scattering is only of order $\sim$0.1~$\mu$m. In comparison, their scattering estimates indicate that a $\sim$15--20~$\mu$m tape-based PM would contribute an emittance growth of order $\sim$10~$\mu$m.\cite{Zingale2021,Barber2020}

A number of LC film formation devices were established and studied by teams at The Ohio State University to balance film-quality requirements with repetition-rate capability.\cite{Poole2015Thesis,Poole2016APL,Poole2016,Czapla2022,Czapla2025,Schumacher2017,Zingale2021,ZingalePhD} Their linear slide target inserter (LSTI) produced 4-octyl-4’-cyanobiphenyl (8CB) films of high optical quality with tunable thickness, but its reliable film formation was limited to approximately once every minute,\cite{Schumacher2017, Czapla2022} significantly lower that required for plasma mirror applications at typical high power laser systems operating at $\geq$1 Hz. The angular fluctuation of films formed with the LSTI was reported to be on the order of 1~mrad.\cite{Czapla2022} The spinning disk inserter (SDI) improved upon the repetition-rate limitation (3 Hz)\cite{Czapla2025}; however, because it relied on apertures on a rotating disk, the angular stability of 8CB films wiped with the SDI was significantly worse than that of the LSTI, with reported pointing fluctuations of about 4~mrad.\cite{Schumacher2017,Czapla2025}

Additional studies, performed by a team led by The Ohio State University, have emphasized characterization and optimization of LC PM performance up to petawatt-level laser power, including high-fluence laser redirection and  pulse-contrast improvement.\cite{Zingale2021,Czapla2025} In this context, Czapla \textit{et al.} demonstrated reflection of 20~J, 35~fs laser pulses from a novel double plasma mirror (DPM) based on free-standing LC films\cite{Czapla2025} formed using SDIs. This work achieved a total DPM throughput exceeding 80\% at a temporal contrast improvement of 2-3 orders of magnitude. In addition, measurements showed a  DPM-reflected beam RMS wavefront error of $100 \pm 20$~nm (Peak-to-Valley $510 \pm 79$~nm). The intrinsic wavefront of the laser pulse was approximately 80~nm RMS in this experiment. 

These previous results support the exploration of LC film formation devices as replenishable PMs in high-power laser facilities. However, a systematic study characterizing film formation reliability, optical quality, and angular fluctuation from film to film, at increasing wiping speed rates, is currently missing. Here, we characterized 8CB LC films produced using a windmill device (the principle of which was first developed by The Ohio State University), evaluating their optical quality through wavefront measurements and quantifying film-to-film orientation stability. The windmill employs a stationary film aperture with multiple rotating wipers sweeping by, enabling higher film-replenishing rates while improving angular repeatability compared to previous film formation devices. We further assessed the reliability of film formation over a range of operating conditions that are relevant to high-repetition-rate use. The impact from the presence of a high-power laser pulse on film-formation reliability was investigated.

\begin{figure*}[t!]
    \centering
    \includegraphics[width=\linewidth]{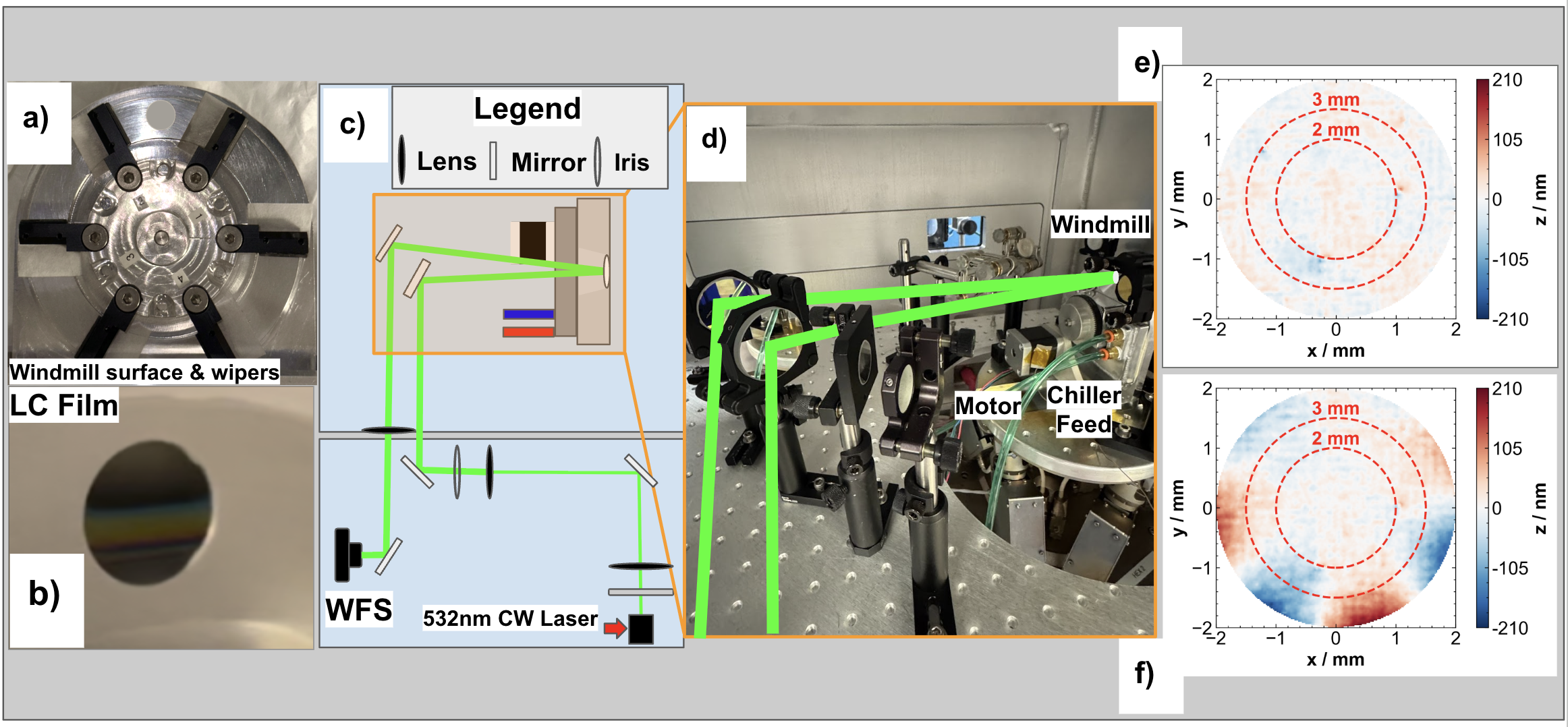}
    \caption{\label{fig} (a) Photograph of the windmill film formation device, including six rotating wipers, the wiping surface, and the 10 mm aperture at the top.
(b) Close-up view of aperture after the liquid crystal (LC) film was wiped. 
(c) Layout of the diagnostic setup illustrating 532 nm continuous-wave (CW) diagnostic beam path, windmill device, and wavefront sensor (WFS). 
(d) Photographic view of the diagnostic setup including the windmill device with motor, chiller cooling cables, and cooling plates. 
(e) Reference wavefront map from a 4 mm diameter region of the gold flat mirror using the same measurement geometry.
(f) Example wavefront map measured over a 4 mm diameter region at the film center after subtracting the reference wavefront depicted in (e).}
\end{figure*}

\section{Liquid Crystal Film Formation and Test Setup}

We investigated LC films, specifically those formed from 4-octyl-4'-cyanobiphenyl (8CB).\cite{Chaban, Singh2000PhaseTransitionsInLiquidCrystals, SinghDunmur2002LiquidCrystalsFundamentals, Poole2016APL, Poole2015Thesis, Schumacher2017, Matsuhashi2007} 8CB has a smectic phase that allows for the formation of free-standing, ultra-thin (10 nm-50 \textmu m) films.\cite{Schumacher2017} Consequently, a small volume of LC can generate thousands of films, providing an inexpensive solution that produces minimal debris.

\begin{figure}[h]
    \centering

    \includegraphics[width=.66\linewidth]{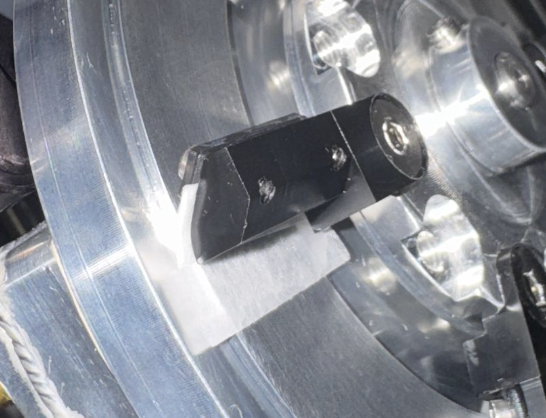}
    \vspace{0.5em}
    \includegraphics[width=.66\linewidth]{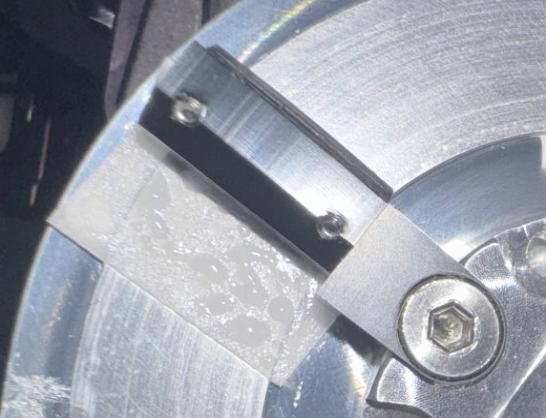}

    \caption{Side view of lens tissue wiper held on windmill surface (top). Front view of lens tissue wiper (bottom).}
    \label{fig:lc_wavefront_ref}
\end{figure}

Maintaining the temperature of the LC in the smectic phase (20-30$^\circ$C\cite{Chaban, Ozgan2011Thermal}) allows the intrinsic surface tension of the PM to produce an optically smooth film surface with nearly uniform film thickness.\cite{Poole2016} If the assembly is operated outside of this smectic phase, it could transition into unusable phases such as: crystalline (<20 $^\circ$C), nematic (approximately 30-40 $^\circ$C), or liquid (>40 $^\circ$C).\cite{Chaban, Ozgan2011Thermal} When operating the assembly in vacuum,  the windmill motor was found to heat the LC to nearly 30$^\circ$C, so a ThermoTek T257P-20 recirculating chiller was connected to a cooling plate attached to the rear of the windmill assembly to mitigate the motor's thermal load. Thermocouple sensors mounted on the windmill surface were used to measure its temperature.

\begin{figure*}[t!]
    \centering
    \includegraphics[width=0.75\linewidth]{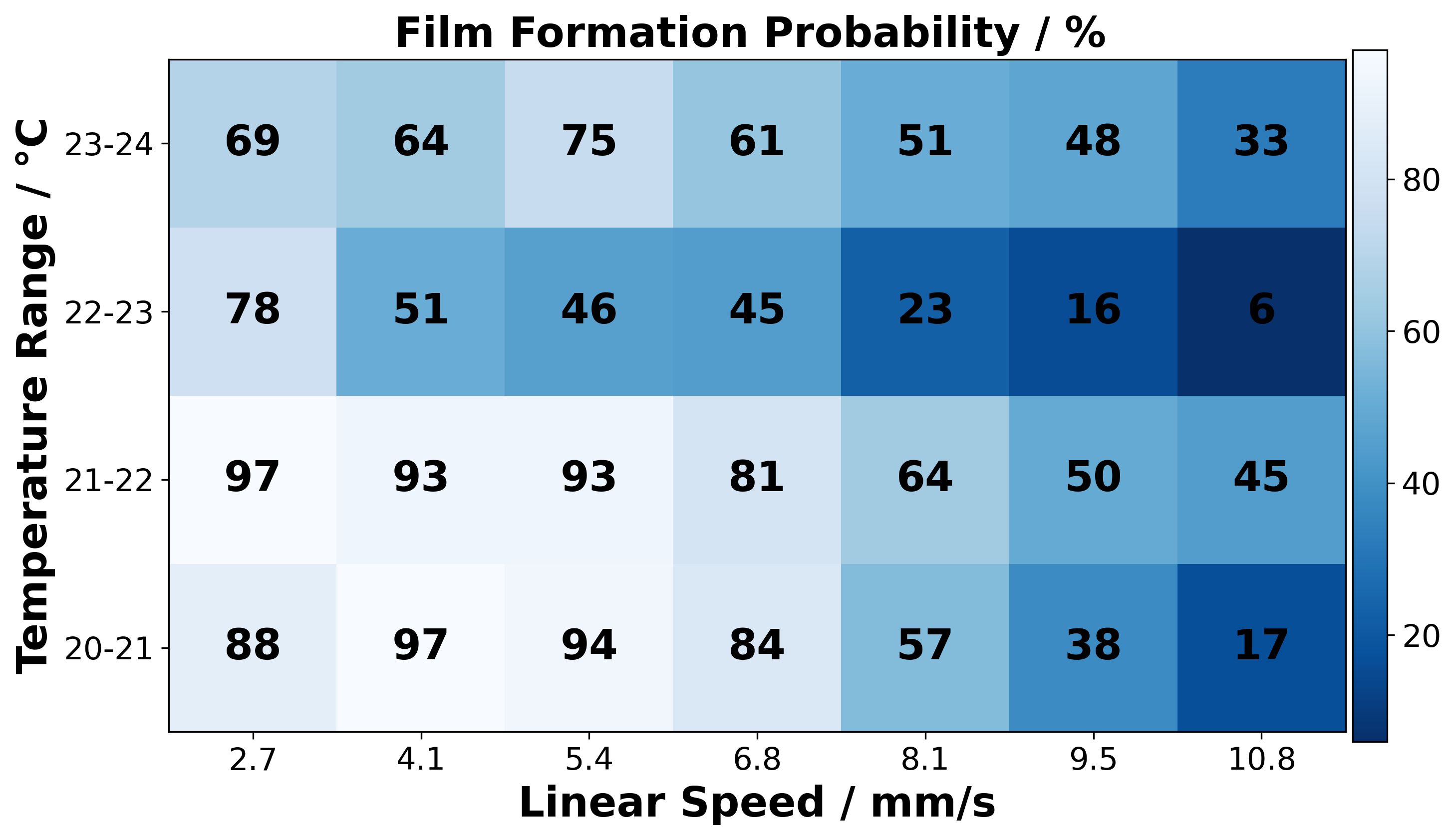}
    \caption{\label{fig}Film formation probability (\%) as a function of windmill surface temperature and wiping speed. For reference, at a linear speed of 2.7 mm/s, the windmill wheel performs a full rotation over the aperture in 100 s, corresponding to a nominal film formation frequency of 0.06 Hz with six wiper arms attached to the wheel, and 0.12 Hz with 12 wiper arms, if every wipe formed a film. At a linear speed of 10.8 mm/s, the film-wiping frequency is increased to 0.24 Hz with 6 arms, or 0.48 Hz for 12 arms.}
\end{figure*}

The front of the windmill is shown in Fig. 1(a). A stepper motor drove a rotor configured with six arms (expandable to twelve), each holding one end of a 10~mm $\times$ 12~mm wiper (length $\times$ width, a close-up is shown in Fig. 2) that traversed the stationary 10 mm diameter aperture on a polished aluminum plate. The wiper was made of lens cleaning tissue purchased from Thorlabs, folded once and mounted in such a way to ensure flat contact with the aluminum surface. This configuration facilitated a consistent wiping mechanism while avoiding damage to the polished surface. To ensure a reproducible LC reservoir on the surface after new LC was applied, a conditioning protocol was performed where the motor was run at a slow speed for 10--15 minutes to evenly distribute the material. Excess liquid was then manually removed from the aperture edge and wiper tissues to prevent accumulation at the perimeter, which would otherwise degrade the reflected wavefront. Following this wiping process, the resulting films (Fig. 1(b)) were found to maintain a central flat region of approximately 3--4~mm across the 10~mm aperture, as confirmed through diagnostic testing described in this paper.

Optical characterization of LC films in vacuum was conducted using a low power diagnostic laser beam as illustrated in the schematic layout in Fig. 1(c). A 532~nm continuous-wave (CW) diode laser was used for wavefront measurements with an Imagine Optic HASO LIFT 680 wavefront sensor.

As shown in Fig. 1(c) and 1(d), the diagnostic beam was reflected from films on the side of the windmill assembly that is opposite of the side containing the wipers. The beam was incident on the surface at $14^{\circ}$ to the normal and was slightly larger than the aperture size to allow full-profile film imaging. Before reaching the wavefront sensor, the reflected beam traversed a 200 mm focal-length lens that imaged the film plane to the wavefront sensor chip. With this imaging relay, the film-to-sensor magnification was $M\approx \num{1.14}$, and the smallest reliably resolvable spatial period on the LC film was 35 \textmu m. The wavefront sensor had an absolute wavefront measurement uncertainty of $\lambda/100$ RMS (approximately $\SI{5.3}{nm}$ RMS at $\lambda=\SI{532}{nm}$).

To accurately assess the film's intrinsic optical quality over a multi-mm central area, the aberrations introduced by the LC film on the reflected diagnostic beam had to be isolated from the beam's intrinsic wavefront. For this purpose, reference wavefronts (see Fig. 1(e)) were recorded by replacing the film with a 1~inch diameter gold mirror with a manufacturer-specified surface flatness of $\lambda/10$ peak-to-valley at 633~nm. Independent measurements indicated that the mirror flatness over a central 4 mm diameter region was $< \SI{5}{nm}$ (the noise floor of the wavefront sensor). The optical aberrations attributable to the LC film were isolated by subtracting the reference wavefront from the corresponding film wavefront measurement. Fig. 1(f) displays a representative film wavefront measurement after subtracting the reference wavefront (z-axis corresponds to the out-of-plane wavefront displacement relative to the reference plane at each transverse position). In this example, the lowest-order Zernike coefficients (piston, tip, tilt, and defocus) were computationally removed to isolate film-induced higher-order aberrations. The defocus term was consistently measured at or below the noise floor of the wavefront sensor.
The angular stability of films formed with the windmill was derived from the tip/tilt Zernike coefficients of the reference-subtracted wavefront data.

\begin{figure*}[t!]
    \centering

    \begin{subfigure}[t]{0.33\linewidth}
        \centering
        \includegraphics[width=\linewidth]{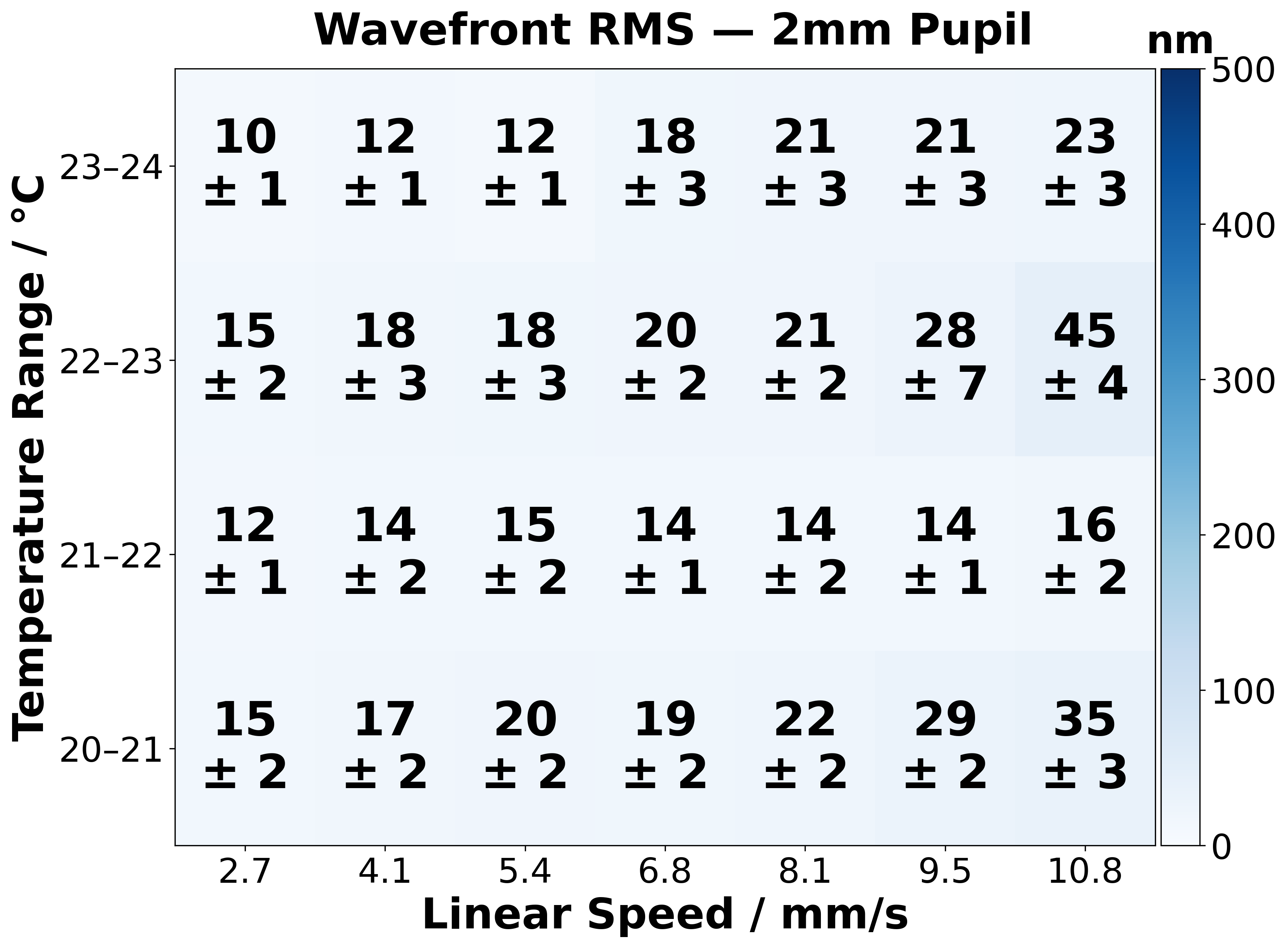}
        \caption{}
        \label{fig:sub1}
    \end{subfigure}\hfill
    \begin{subfigure}[t]{0.33\linewidth}
        \centering
        \includegraphics[width=\linewidth]{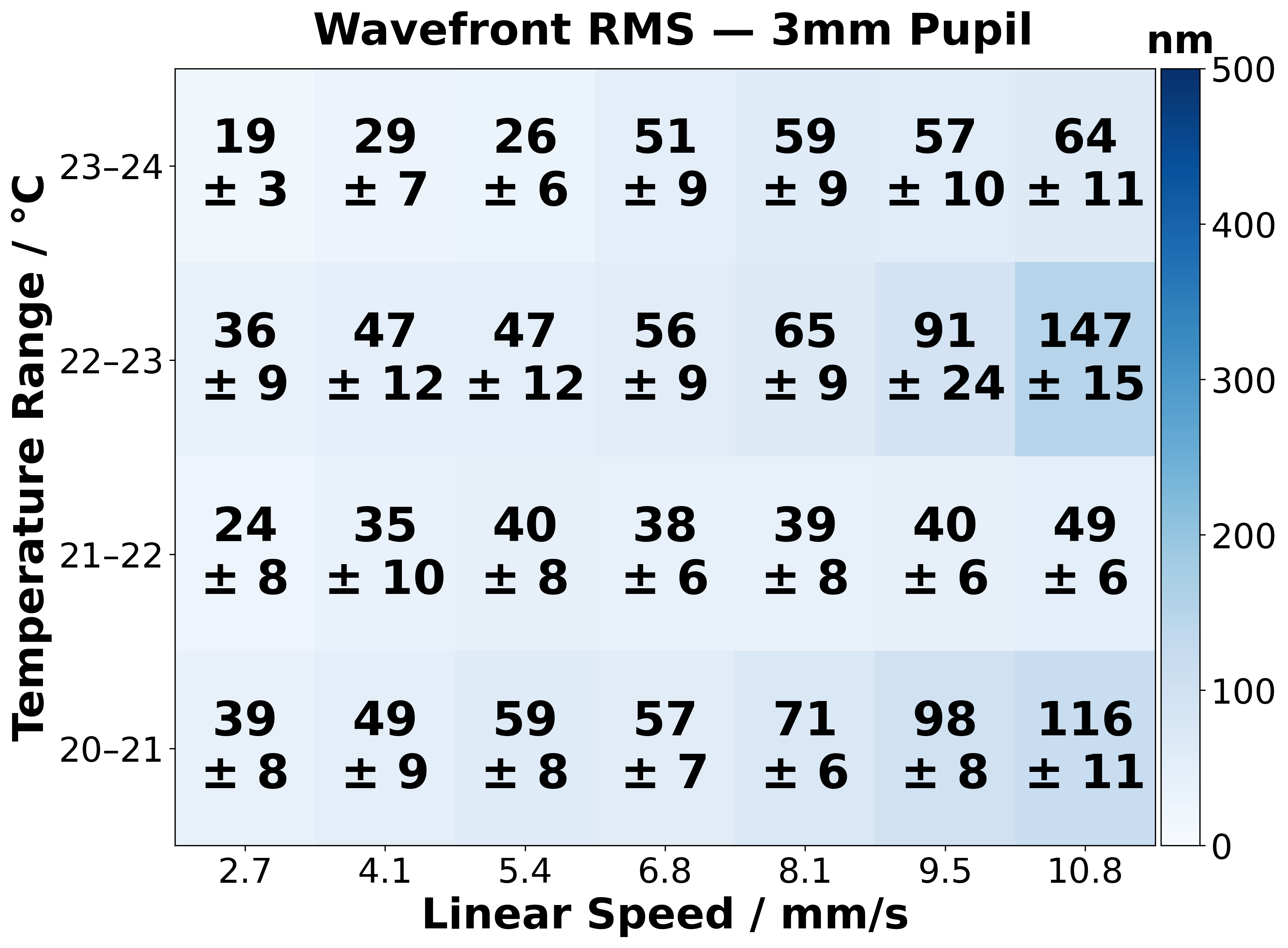}
        \caption{}
        \label{fig:sub2}
    \end{subfigure}\hfill
    \begin{subfigure}[t]{0.33\linewidth}
        \centering
        \includegraphics[width=\linewidth]{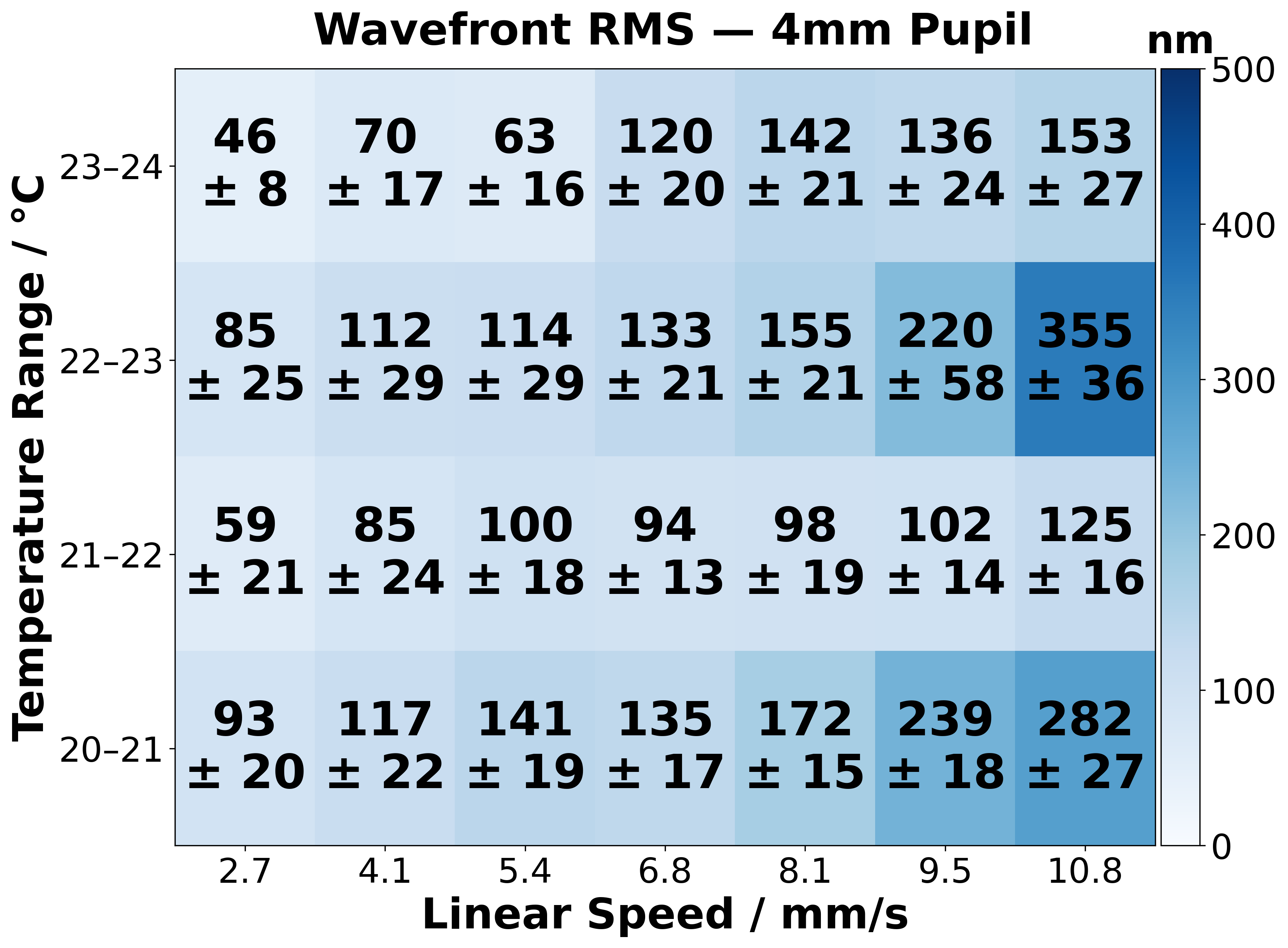}
        \caption{}
        \label{fig:sub3}
    \end{subfigure}
    \caption{Heatmaps summarizing the optical quality of 8CB liquid crystal films as a function of windmill surface temperature and wiping speed. Here, 2.7 mm/s corresponds to a nominal film exposure frequency of 0.06 Hz using 6 arms, and 0.12 Hz using 12 arms; 10.8 mm/s corresponds to a nominal film exposure frequency of 0.24 Hz for 6 arms, or 0.48 Hz for 12 arms. For analysis pupil diameters of (a) 2~mm, (b) 3~mm, and (c) 4~mm, the wavefront root-mean-square (RMS) values (nm) after subtraction of a flat-mirror reference, where in each cell, the displayed value corresponds to the mean $\pm$ standard deviation over 120 film-formation attempts in vacuum per grid configuration, excluding piston, tip, tilt, and defocus Zernike terms for this analysis.}
    \label{fig:combined_plots}
\end{figure*}

\section{Characterization of Optical Quality, Angular Stability and Formation Reliability of films produced with the Windmill}

We characterized the reflected wavefront, film-to-film angular stability, and reliability of films formed with the windmill device, across a range of film wiping speeds and assembly temperatures. The impact on the wavefront of beams of different diameters was assessed by applying varying analysis pupil sizes before calculating the wavefront RMS. Circles in Figs. 1(e) and 1(f) indicate the pupil sizes used in this study (2, 3, and 4 mm). While we measure the full 10 mm aperture to capture and analyze edge effects, we display the 4 mm sub-aperture which corresponds to the aperture a high-power beam would be incident on.

Before presenting wavefront and angular fluctuation results, we first quantify the reliability of 8CB film formation at increasing wiper speeds to determine the maximum repetition rate. Using a six arm configuration, we measured the success rate of films being formed over 120 attempts to wipe a film for each wiping speed. The wiping speed is the linear speed of each wiper as it traverses over the film aperture. A successful wipe was defined by the appearance of a reflected beam on our wavefront sensor, while a failure was logged if no beam was observed after the wiper passed the aperture. The resulting formation probabilities with respect to wiper speed are presented in Fig. 3 and span from 6\% at 10.8 mm/s to 97\% at 2.7 mm/s. 2.7 mm/s translates to a nominal film wiping frequency of 0.06 Hz for the six-arm device used, and 0.12 Hz if the number of arms was increased to 12 which is the full capacity of the current windmill device; 10.8 mm/s would translate to a nominal film wiping frequency of 0.24 Hz for 6 arms, or 0.48 Hz for 12 arms. The effective film formation frequency is calculated using \(f_{\mathrm{eff}}(T,v)=f_0(v)\,P(T,v)/100\), where \(f_0\) is the film wiping frequency and \(P\) is the film formation probability. As visible in Fig. 3, the film formation probability depended on the interplay between wiping speed and assembly temperature. Within the optimal 21--22~$^{\circ}$C range, the windmill device achieved high reliability ($>90\%$) for all film formation speeds up to 5.4 mm/s. With a formation reliability of $84\%$, the highest effective film formation frequency was observed for a wiping speed of 6.8~mm/s within 20--21~$^{\circ}$C, resulting in approximately 0.13~Hz for the six-arm device and 0.25~Hz for 12 arms. However, as the wiping speed approached the 11 mm/s regime, the formation probability decreased significantly, falling to 50\% at 9.5 mm/s and below 10\% at certain higher speed/temperature combinations.

We suspect that this formation probability trend arises from coupled physical limitations at the moving LC boundary: (i) insufficient relaxation time for smectic ordering to re-establish after rapid drawing, and (ii) stress/defect-mediated interfacial dynamics localized near the aperture edge. In smectic materials, ordering is governed by reorientation/reorganization kinetics that depend on how effectively stress can be dissipated during interfacial motion.\cite{rupture2014} As the film edge is a natural site for elevated stress concentration, it may dominate defect creation and evolution during boundary motion.\cite{Stannarius2019}
This molecular instability may also be modulated by the microscopic texture of the aperture edge. Liquid crystals can form boundary-stabilized configurations by adapting their molecular arrangement to surface features.\cite{Abbott1997} Such behavior is consistent with roughness-dependent effective anchoring/pinning and a preferred configuration that minimizes interfacial stress.\cite{Jerome1991} Therefore, the edge roughness likely governs attachment robustness during rapid wiping: if too smooth, insufficient anchoring/pinning may destabilize the configuration and promote rupture, whereas an overly jagged surface can disrupt near-surface order and enhance defect formation.\cite{Jerome1991}

The optical quality of the LC films across different pupil sizes, temperatures, and wiping speeds using said six arm configuration are summarized in Fig. 4. Across all tested conditions, the wavefront RMS of higher order Zernike terms (excluding piston, tip, tilt, and defocus Zernike terms) for the 2~mm diagnostic pupil ranged from 10 to 45~nm, with corresponding peak-to-valley (PV) values between 83 and 272~nm (Fig.~4(a)). We found that the optical quality was most robust within the 21--22$^{\circ}$C temperature window, where the wavefront RMS only varied between 12~nm and 16~nm across all investigated wiping speeds. For a 3~mm diameter pupil (Fig.~4(b)), the average wavefront RMS for the 21--22$^{\circ}$C temperature range was between 24~nm and 49~nm, with peak RMS values reaching up to 147~nm across all conditions. For a 4~mm diameter pupil (Fig.~4(c)), the average RMS for the 21--22$^{\circ}$C range spanned 59~nm to 125~nm, with the maximum aberrations reaching 355~nm RMS; the corresponding PV values are provided in the supplementary material. Because film formation became less reliable at higher wiping frequencies, fewer wipe attempts successfully formed a film. As a result, the number of data points available for the statistical analysis in Fig. 4 decreased at higher wiping frequencies.

\begin{figure}[h]
    \centering

    \begin{subfigure}[t]{.99\linewidth}
        \centering
        \includegraphics[width=\linewidth]{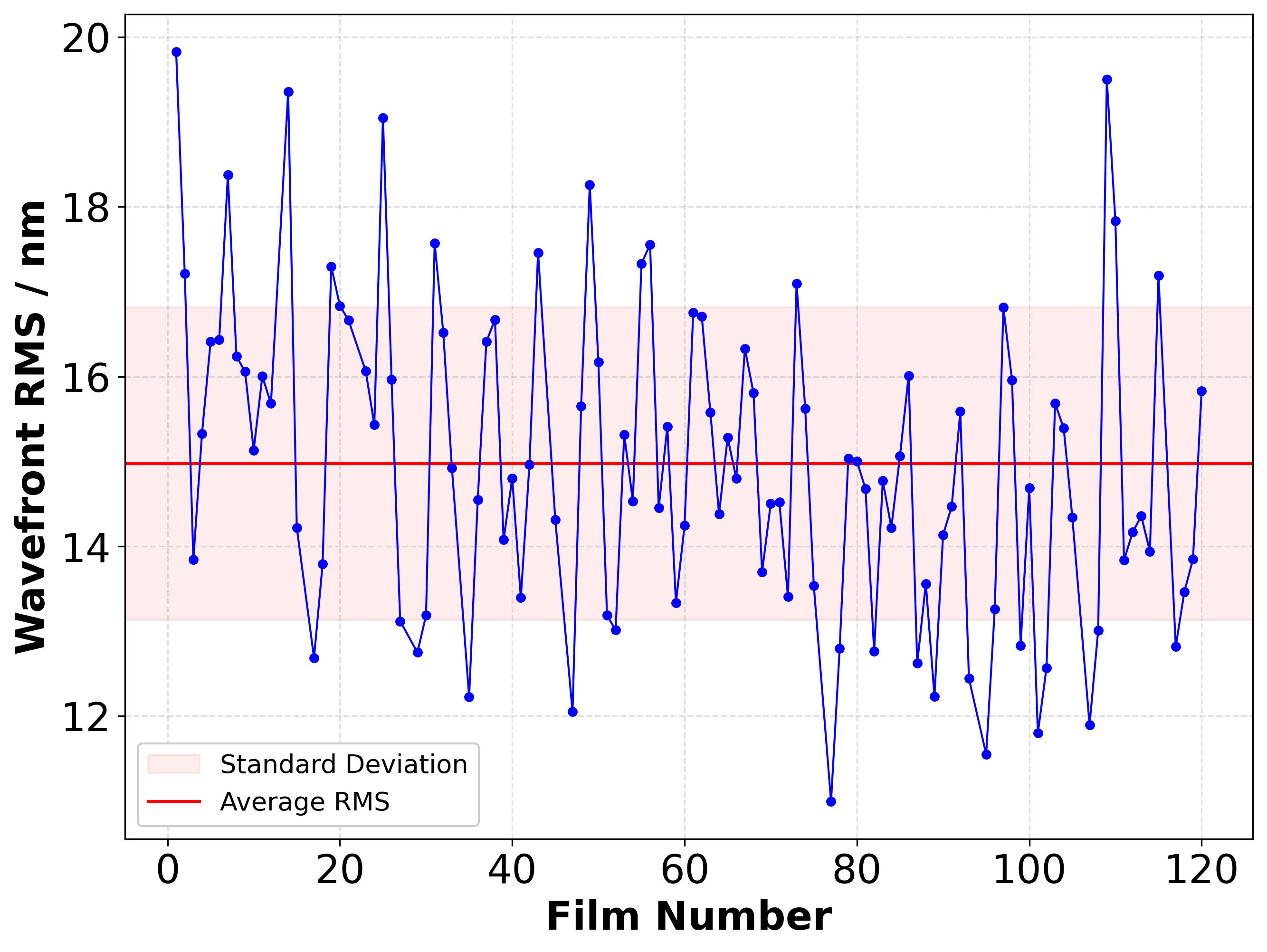}
        \caption{}
        \label{fig:lc_wavefront_rms_pv_2122}
    \end{subfigure}

    \vspace{0.01em}

    \begin{subfigure}[t]{.99\linewidth}
        \centering
        \includegraphics[width=\linewidth]{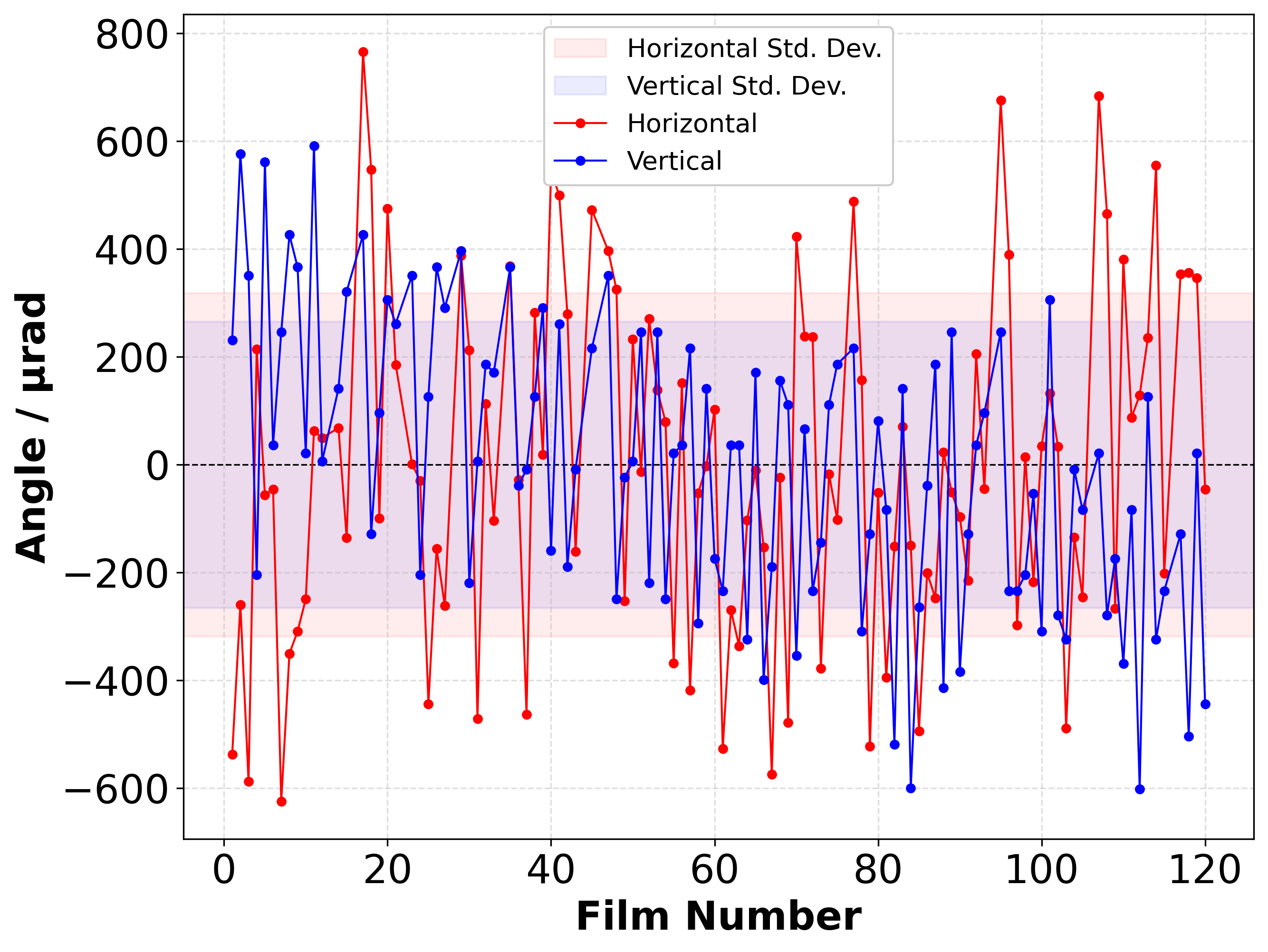}
        \caption{}
        \label{fig:lc_pointing_2122}
    \end{subfigure}

    \caption{Panel (a): Example wavefront RMS fluctuation for a windmill surface temperature of 21-22 °C and wiping speed of 5.4 mm/s (corresponding to an effective film formation frequency of 0.11 Hz and 0.22 Hz for 6 and 12 wipers, respectively). Panel (b): Horizontal and vertical angular fluctuation under the same conditions.}
    \label{fig:lc_wavefront_ref}
\end{figure}

Across the full range of temperatures and wiping speeds, the total angular fluctuation ($\Delta\theta_{\text{total}}$) spanned from approximately 270 to 720~\textmu rad, which is within the range of standard motorized actuators for tip/tilt correction (all film angle measurements are listed in the supplementary material). 

Film-to-film fluctuations of the wavefront RMS, analyzed for a 2~mm pupil size, are displayed in Fig.~5(a) for an operating condition of 21--22$^{\circ}$C and 5.4 mm/s wiping speed. The corresponding film-to-film angular fluctuations for the same operating conditions are shown in Fig.~5(b). The horizontal angular fluctuation exceeded the vertical component, a trend that was observed across nearly all tested configurations. This asymmetry indicated that the horizontal wiping direction of the windmill arms was consistently the less stable axis for beam pointing. Additionally, both wavefront RMS and angular pointing measurements of individual subsequent films revealed periodic film-to-film trends. A possible explanation is that adjacent wipers cut from the same tissue sheet were not perfectly identical: individual wipers could therefore generate films with distinct aberration signatures and corresponding repeatable angle deviations. This suggests that further standardization of wiper production could reduce film-to-film variations of optical quality. In addition, we observed slow drifts when wipers were not exchanged on the order of a week or when operating at higher temperatures/speeds. We attributed this behavior to excess LC accumulating unevenly at the trailing edge of the aperture during horizontal wiping. This could have perturbed the evolving meniscus profile, producing a correlated angular drift and reflected wavefront degradation.

\section{HIGH-POWER DEMONSTRATION OF WINDMILL LC FILM FORMATION AT THE BELLA PETAWATT LASER}

In this work, the BELLA PW laser\cite{Nakamura2017} at LBNL was used to test windmill-formed 8CB LC films as replenishable PMs capable of redirecting high-power pulses into a beam dump to protect downstream components. The transmitted laser energy was measured through the LC film, while the reflected pulse was directed into a beam dump 20 cm away. In this configuration, reduced transmission relative to shots without a film in place indicated plasma formation at the LC surface by the pulse's leading edge, which redirected a fraction of the incident energy away from the transmission diagnostic and toward the dump, thereby protecting downstream hardware. This measurement also served to establish that the windmill assembly was able to perform under the extreme conditions characteristic of the plasma mirror environment, including direct and diffuse laser irradiation, plasma expansion, and debris production.

The windmill was installed at a location 19 cm downstream of the laser focus, where the laser beam diameter (FWHM) on the films was approximately $w_x = 2.1$ mm (horizontal) and $w_y = 2.0$ mm (vertical), using a standard F/65 off-axis parabolic mirror.\cite{Nakamura2017} As we increased the laser pulse energy from 0.7 J to 8.8 J, the incident fluence on the films ranged from $16~\mathrm{J/cm^2}$ to $215~\mathrm{J/cm^2}$. The reported fluences and beam FWHM are interpolated from high-power laser mode measurements taken $\sim$1~cm downstream and $\sim$1.5~cm upstream of the windmill location, and are given at the film surface, accounting for the 26\textdegree{} tilt of the windmill from normal incidence, which was used to direct the beam toward the beam dump. For intensity conversion, we used a recent measurement of the temporal laser pulse shape, providing an energy-normalized peak power of $P_E=24~\mathrm{TW/J}$, for the 36 fs laser pulse. We used $P_E$ to convert the maximum applied fluence to intensity as $I \approx \left(215~\mathrm{J/cm^2}\right)\left(24~\mathrm{TW/J}\right) = 5.16\times10^{15}~\mathrm{W/cm^2}$. For the incident fluence range studied here, the corresponding intensities were thus $\sim 3.84\times10^{14}$--$5.16\times10^{15}~\mathrm{W/cm^2}$. At these intensities, the LC is expected to be rapidly ionized by the leading edge of the pulse, forming a plasma mirror. The transmitted light through the plasma mirror was diagnosed using a ceramic screen located behind a wedge downstream of the interaction, monitoring the laser spatial mode.\cite{Gonsalves2019Petawatt, Nakamura2017, picksley2024} The ceramic screen was imaged to a diagnostic camera using a lens.

\begin{figure}[h]
    \centering
    \includegraphics[width=\linewidth]{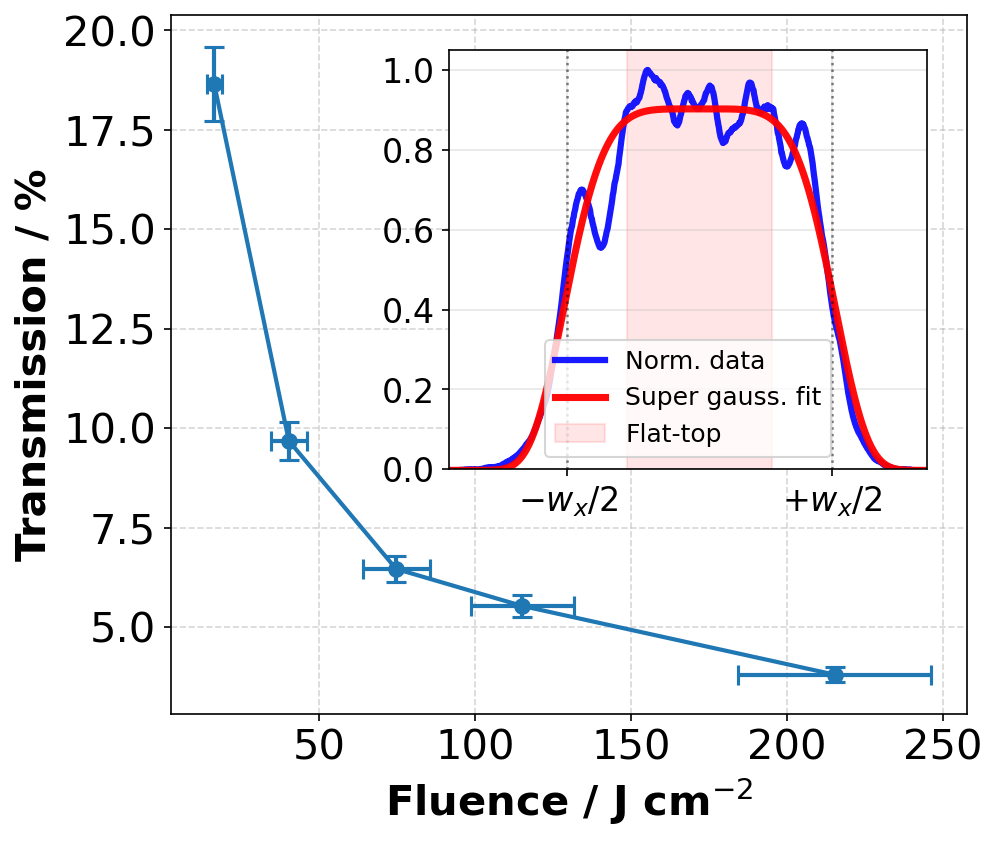}
    \caption{\label{fig} Measured transmission as a function of laser fluence for free-standing 8CB liquid crystal (LC) films, using the BELLA PW laser at selected pulse energies. Liquid crystal films were produced across a 10~mm aperture using the windmill device, and exposed to laser pulses with an incident beam diameter of $w_x = 2.1$~mm and $w_y = 2.0$~mm (full width at half maximum, FWHM) at the film plane. The inset shows a normalized horizontal lineout of the incident beam profile on the films, measured $\approx$1~cm downstream of the LC film, along with a super-Gaussian fit. Dotted lines indicate the FWHM. The transmission error bars include the laser pulse energy fluctuation shot-to-shot (standard deviation) and the measurement uncertainty of the transmission diagnostic. The fluence error bars include uncertainties of the energy fluctuation shot-to-shot (standard deviation) and the fluence fluctuation (standard deviation) across the flat-top region of the beam profile incident on the films (within red shaded region of the inset).}
\end{figure}

The results of the transmission measurements are displayed in Fig.~6. Reference measurements were collected by propagating the laser pulse through the center of the 10~mm windmill aperture under identical conditions but with no LC film present. The transmission percentage was then determined by the ratio of the mean counts on the diagnostic camera after background subtraction for shots with a film relative to without a film. Each transmission data point in Fig.~6 represents an average over three laser shots on individual films. At the lowest fluence, $16 \text{ J/cm$^2$}$, the measured transmission was (18 $\pm$ 1)\%. This value is substantially lower than the expected low-power transmission ($>85\%$) for these few-10~nm-thick films in the absence of a plasma.\cite{Poole2016} As the incident energy was increased, a clear decrease in transmission was observed, consistent with plasma formation on the film driven by the leading edge of the same laser pulse. In this scenario, a portion of the pulse is reflected by the transient plasma and redirected away from the transmission diagnostic toward the dump. At the maximum fluence, $215 \text{ J/cm$^2$}$, the transmission reached a minimum of (4.0 $\pm$ 0.2)\%. Previously reported PM reflection measurements showed a pronounced increase over the same fluence range as studied here.\cite{Shaw2016,Doumy2004PlasmaMirror} Although direct reflectivity measurements were not recorded—since the LC PM is used primarily to protect downstream components—the measured transmission behavior hints to a similar trend. For reference, when used to redirect the stage 1 drive laser away from downstream components in the multi-GeV staging experiment at BELLA, a fluence on the LC film of $F_{\mathrm{stage}} \simeq 415~\mathrm{J/cm^2}$ (corresponding to an intensity $I \approx 1\times10^{16}~\mathrm{W/cm^2}$) is anticipated. In between laser shots, each wiping attempt at 1.35~mm/s produced a film, consistent with the film formation reliability results from the low-power characterizations shown in Fig.~3; moreover, the windmill maintained 100\% formation reliability within the $21$--$22^\circ\mathrm{C}$ temperature window, including after the high-power campaign. No noticeable amounts of debris were observed in the target chamber after completing these 15 test shots.

\section{Conclusion}
This work establishes windmill-formed 8CB LC films as a replenishable PM technology for high-power laser applications, with a maximum effective film formation frequency of \textasciitilde0.25~Hz. We observed the highest film quality when operating in a temperature range of $21$--$22^\circ\mathrm{C}$, in which the films added a wavefront RMS error of only 12--16~nm over a 2~mm analysis pupil, 24--49~nm over a 3~mm pupil, and 59--125~nm over a 4~mm pupil. Furthermore, film-to-film angular pointing fluctuations were constrained to $\le 0.52$~mrad, and the device maintained $>$90\% film formation reliability for wiping speeds up to 5.4~mm/s. Additionally, we demonstrated robust film formation and laser energy redirection during high-power tests using the BELLA PW laser. Future work will focus on further reducing residual pointing jitter through active stabilization, improving film-formation reliability at higher repetition rates by standardizing wiper production, testing alternative wiper geometries, surface treatments to enhance windmill aperture adhesion, and standardized LC volume application to the tissues, aiming to extend reliable operation toward the $\sim$1~Hz regime required for full-rate deployment in next-generation high-power laser applications.

\section*{Data availability}
The data that support the findings of this study are openly available in Zenodo at \href{http://doi.org/10.5281/zenodo.17138268}{doi.org/10.5281/zenodo.17138268}, reference number [\cite{repo}].

\section*{Supplementary Material}
See the supplementary material for wavefront RMS and peak-to-valley statistics and film pointing angles as a function of temperature range, analysis pupil size, and linear speed.

\section*{Acknowledgements}
This work was supported by U.S. Department of Energy Office of Science, Office of High Energy Physics under Contract No. DE-AC02-05CH11231. We greatly acknowledge technical support by Mark Kirkpatrick, Teo Maldonado Mancuso, and Derrick McGrew, and software support by Chetanya Jain. We would also like to thank Sam Barber, Joshua Stackhouse, and Raymond Li for useful discussions.

\bibliography{aipsamp_v2_unbolded}

\providecommand{\noopsort}[1]{}\providecommand{\singleletter}[1]{#1}%
\begin{thebibliography}{39}%
\makeatletter
\providecommand \@ifxundefined [1]{%
 \@ifx{#1\undefined}
}%
\providecommand \@ifnum [1]{%
 \ifnum #1\expandafter \@firstoftwo
 \else \expandafter \@secondoftwo
 \fi
}%
\providecommand \@ifx [1]{%
 \ifx #1\expandafter \@firstoftwo
 \else \expandafter \@secondoftwo
 \fi
}%
\providecommand \natexlab [1]{#1}%
\providecommand \enquote  [1]{``#1''}%
\providecommand \bibnamefont  [1]{#1}%
\providecommand \bibfnamefont [1]{#1}%
\providecommand \citenamefont [1]{#1}%
\providecommand \href@noop [0]{\@secondoftwo}%
\providecommand \href [0]{\begingroup \@sanitize@url \@href}%
\providecommand \@href[1]{\@@startlink{#1}\@@href}%
\providecommand \@@href[1]{\endgroup#1\@@endlink}%
\providecommand \@sanitize@url [0]{\catcode `\\12\catcode `\$12\catcode `\&12\catcode `\#12\catcode `\^12\catcode `\_12\catcode `\%12\relax}%
\providecommand \@@startlink[1]{}%
\providecommand \@@endlink[0]{}%
\providecommand \url  [0]{\begingroup\@sanitize@url \@url }%
\providecommand \@url [1]{\endgroup\@href {#1}{\urlprefix }}%
\providecommand \urlprefix  [0]{URL }%
\providecommand \Eprint [0]{\href }%
\providecommand \doibase [0]{http://dx.doi.org/}%
\providecommand \selectlanguage [0]{\@gobble}%
\providecommand \bibinfo  [0]{\@secondoftwo}%
\providecommand \bibfield  [0]{\@secondoftwo}%
\providecommand \translation [1]{[#1]}%
\providecommand \BibitemOpen [0]{}%
\providecommand \bibitemStop [0]{}%
\providecommand \bibitemNoStop [0]{.\EOS\space}%
\providecommand \EOS [0]{\spacefactor3000\relax}%
\providecommand \BibitemShut  [1]{\csname bibitem#1\endcsname}%
\let\auto@bib@innerbib\@empty
\bibitem [{\citenamefont {Danson}(2019)}]{Danson2019}%
  \BibitemOpen
  \bibfield  {author} {\bibinfo {author} {\bibfnamefont {C.~N.}\ \bibnamefont {Danson}},\ }\bibfield  {title} {\enquote {\bibinfo {title} {Petawatt and exawatt class lasers worldwide},}\ }\href {\doibase 10.1017/hpl.2019.36} {\bibfield  {journal} {\bibinfo  {journal} {High Power Laser Science and Engineering}\ }\textbf {\bibinfo {volume} {7}},\ \bibinfo {pages} {e54} (\bibinfo {year} {2019})}\BibitemShut {NoStop}%
\bibitem [{\citenamefont {Yoon}(2021)}]{Yoon2021}%
  \BibitemOpen
  \bibfield  {author} {\bibinfo {author} {\bibfnamefont {J.~W.}\ \bibnamefont {Yoon}},\ }\bibfield  {title} {\enquote {\bibinfo {title} {Realization of laser intensity over > {$10^{23}$} {W/cm$^2$}},}\ }\href {\doibase 10.1364/OPTICA.420520} {\bibfield  {journal} {\bibinfo  {journal} {Optica}\ }\textbf {\bibinfo {volume} {8}},\ \bibinfo {pages} {630--635} (\bibinfo {year} {2021})}\BibitemShut {NoStop}%
\bibitem [{\citenamefont {Thaury}(2007)}]{Thaury2007}%
  \BibitemOpen
  \bibfield  {author} {\bibinfo {author} {\bibfnamefont {C.}~\bibnamefont {Thaury}},\ }\bibfield  {title} {\enquote {\bibinfo {title} {Plasma mirrors for ultrahigh-intensity optics},}\ }\href@noop {} {\bibfield  {journal} {\bibinfo  {journal} {Nature Physics}\ }\textbf {\bibinfo {volume} {3}},\ \bibinfo {pages} {424--429} (\bibinfo {year} {2007})}\BibitemShut {NoStop}%
\bibitem [{\citenamefont {Sokollik}(2010)}]{Sokollik2010}%
  \BibitemOpen
  \bibfield  {author} {\bibinfo {author} {\bibfnamefont {T.}~\bibnamefont {Sokollik}},\ }\bibfield  {title} {\enquote {\bibinfo {title} {Tape-drive based plasma mirror},}\ }in\ \href {\doibase 10.1063/1.3520330} {\emph {\bibinfo {booktitle} {AIP Conference Proceedings}}},\ Vol.\ \bibinfo {volume} {1299}\ (\bibinfo {year} {2010})\ pp.\ \bibinfo {pages} {233--237}\BibitemShut {NoStop}%
\bibitem [{\citenamefont {Dromey}(2004)}]{Dromey2004PlasmaMirror}%
  \BibitemOpen
  \bibfield  {author} {\bibinfo {author} {\bibfnamefont {B.}~\bibnamefont {Dromey}},\ }\bibfield  {title} {\enquote {\bibinfo {title} {The plasma mirror---a subpicosecond optical switch for ultrahigh power lasers},}\ }\href {\doibase 10.1063/1.1646737} {\bibfield  {journal} {\bibinfo  {journal} {Review of Scientific Instruments}\ }\textbf {\bibinfo {volume} {75}},\ \bibinfo {pages} {645--649} (\bibinfo {year} {2004})}\BibitemShut {NoStop}%
\bibitem [{\citenamefont {Obst}(2018)}]{Obst2018}%
  \BibitemOpen
  \bibfield  {author} {\bibinfo {author} {\bibfnamefont {L.}~\bibnamefont {Obst}},\ }\bibfield  {title} {\enquote {\bibinfo {title} {{On-shot} characterization of single plasma mirror temporal contrast improvement},}\ }\href {\doibase 10.1088/1361-6587/aab3bb} {\bibfield  {journal} {\bibinfo  {journal} {Plasma Physics and Controlled Fusion}\ }\textbf {\bibinfo {volume} {60}},\ \bibinfo {pages} {054007} (\bibinfo {year} {2018})}\BibitemShut {NoStop}%
\bibitem [{\citenamefont {Shaw}(2016)}]{Shaw2016}%
  \BibitemOpen
  \bibfield  {author} {\bibinfo {author} {\bibfnamefont {B.~H.}\ \bibnamefont {Shaw}},\ }\bibfield  {title} {\enquote {\bibinfo {title} {Reflectance characterization of tape-based plasma mirrors},}\ }\href {\doibase 10.1063/1.4954242} {\bibfield  {journal} {\bibinfo  {journal} {Phys. Plasmas}\ }\textbf {\bibinfo {volume} {23}},\ \bibinfo {pages} {063118} (\bibinfo {year} {2016})}\BibitemShut {NoStop}%
\bibitem [{\citenamefont {Gruse}(2025)}]{Gruse2025SelfGuidedPlasmaMirrors}%
  \BibitemOpen
  \bibfield  {author} {\bibinfo {author} {\bibfnamefont {J.-N.}\ \bibnamefont {Gruse}},\ }\bibfield  {title} {\enquote {\bibinfo {title} {Self-guided propagation of laser pulses reflected at high intensity from plasma mirrors},}\ }\href {\doibase 10.1088/1367-2630/ae2cbf} {\bibfield  {journal} {\bibinfo  {journal} {New Journal of Physics}\ }\textbf {\bibinfo {volume} {27}},\ \bibinfo {pages} {124302} (\bibinfo {year} {2025})}\BibitemShut {NoStop}%
\bibitem [{\citenamefont {Prencipe}(2017)}]{Prencipe2017TargetsHighRepetitionRate}%
  \BibitemOpen
  \bibfield  {author} {\bibinfo {author} {\bibfnamefont {I.}~\bibnamefont {Prencipe}},\ }\bibfield  {title} {\enquote {\bibinfo {title} {Targets for high repetition rate laser facilities: needs, challenges and perspectives},}\ }\href {\doibase 10.1017/hpl.2017.18} {\bibfield  {journal} {\bibinfo  {journal} {High Power Laser Science and Engineering}\ }\textbf {\bibinfo {volume} {5}},\ \bibinfo {pages} {e17} (\bibinfo {year} {2017})}\BibitemShut {NoStop}%
\bibitem [{\citenamefont {Barber}(2020)}]{Barber2020}%
  \BibitemOpen
  \bibfield  {author} {\bibinfo {author} {\bibfnamefont {S.~K.}\ \bibnamefont {Barber}},\ }\bibfield  {title} {\enquote {\bibinfo {title} {A compact, high resolution energy, and emittance diagnostic for electron beams using active plasma lenses},}\ }\href {\doibase 10.1063/5.0005114} {\bibfield  {journal} {\bibinfo  {journal} {Applied Physics Letters}\ }\textbf {\bibinfo {volume} {116}},\ \bibinfo {pages} {234108} (\bibinfo {year} {2020})}\BibitemShut {NoStop}%
\bibitem [{\citenamefont {Steinke}(2016{\natexlab{a}})}]{steinke2016}%
  \BibitemOpen
  \bibfield  {author} {\bibinfo {author} {\bibfnamefont {S.}~\bibnamefont {Steinke}},\ }\bibfield  {title} {\enquote {\bibinfo {title} {Multistage coupling of independent laser-plasma accelerators},}\ }\href {\doibase 10.1038/nature16525} {\bibfield  {journal} {\bibinfo  {journal} {Nature}\ }\textbf {\bibinfo {volume} {530}},\ \bibinfo {pages} {190--193} (\bibinfo {year} {2016}{\natexlab{a}})}\BibitemShut {NoStop}%
\bibitem [{\citenamefont {Steinke}(2016{\natexlab{b}})}]{steinke2016staging}%
  \BibitemOpen
  \bibfield  {author} {\bibinfo {author} {\bibfnamefont {S.}~\bibnamefont {Steinke}},\ }\bibfield  {title} {\enquote {\bibinfo {title} {Staging of laser-plasma accelerators},}\ }\href@noop {} {\bibfield  {journal} {\bibinfo  {journal} {Physics of Plasmas}\ }\textbf {\bibinfo {volume} {23}} (\bibinfo {year} {2016}{\natexlab{b}})}\BibitemShut {NoStop}%
\bibitem [{\citenamefont {Schroeder}(2023)}]{schroeder2023linear}%
  \BibitemOpen
  \bibfield  {author} {\bibinfo {author} {\bibfnamefont {C.}~\bibnamefont {Schroeder}},\ }\bibfield  {title} {\enquote {\bibinfo {title} {Linear colliders based on laser-plasma accelerators},}\ }\href@noop {} {\bibfield  {journal} {\bibinfo  {journal} {Journal of Instrumentation}\ }\textbf {\bibinfo {volume} {18}},\ \bibinfo {pages} {T06001} (\bibinfo {year} {2023})}\BibitemShut {NoStop}%
\bibitem [{\citenamefont {Esarey}(2009)}]{Esarey2009}%
  \BibitemOpen
  \bibfield  {author} {\bibinfo {author} {\bibfnamefont {E.}~\bibnamefont {Esarey}},\ }\bibfield  {title} {\enquote {\bibinfo {title} {Physics of laser-driven plasma-based electron accelerators},}\ }\href {\doibase 10.1103/RevModPhys.81.1229} {\bibfield  {journal} {\bibinfo  {journal} {Reviews of Modern Physics}\ }\textbf {\bibinfo {volume} {81}},\ \bibinfo {pages} {1229--1285} (\bibinfo {year} {2009})}\BibitemShut {NoStop}%
\bibitem [{\citenamefont {Picksley}(2022)}]{Picksley2022}%
  \BibitemOpen
  \bibfield  {author} {\bibinfo {author} {\bibfnamefont {A.}~\bibnamefont {Picksley}},\ }\bibfield  {title} {\enquote {\bibinfo {title} {Commissioning of the second beamline upgrade to {BELLA PW}},}\ }in\ \href {\doibase 10.1109/AAC55212.2022.10822932} {\emph {\bibinfo {booktitle} {2022 IEEE Advanced Accelerator Concepts Workshop (AAC)}}}\ (\bibinfo {year} {2022})\ pp.\ \bibinfo {pages} {1--5}\BibitemShut {NoStop}%
\bibitem [{\citenamefont {Nakamura}(2017)}]{Nakamura2017}%
  \BibitemOpen
  \bibfield  {author} {\bibinfo {author} {\bibfnamefont {K.}~\bibnamefont {Nakamura}},\ }\bibfield  {title} {\enquote {\bibinfo {title} {Diagnostics, control and performance parameters for the {BELLA} high repetition rate {Petawatt} class laser},}\ }\href {\doibase 10.1109/JQE.2017.2708601} {\bibfield  {journal} {\bibinfo  {journal} {IEEE Journal of Quantum Electronics}\ }\textbf {\bibinfo {volume} {53}},\ \bibinfo {pages} {1200121} (\bibinfo {year} {2017})}\BibitemShut {NoStop}%
\bibitem [{\citenamefont {Gonsalves}(2019)}]{Gonsalves2019Petawatt}%
  \BibitemOpen
  \bibfield  {author} {\bibinfo {author} {\bibfnamefont {A.~J.}\ \bibnamefont {Gonsalves}},\ }\bibfield  {title} {\enquote {\bibinfo {title} {Petawatt laser guiding and electron beam acceleration to 8 gev in a laser-heated capillary discharge waveguide},}\ }\href {\doibase 10.1103/PhysRevLett.122.084801} {\bibfield  {journal} {\bibinfo  {journal} {Physical Review Letters}\ }\textbf {\bibinfo {volume} {122}},\ \bibinfo {pages} {084801} (\bibinfo {year} {2019})}\BibitemShut {NoStop}%
\bibitem [{\citenamefont {{van Tilborg}}(2015)}]{vanTilborg2015}%
  \BibitemOpen
  \bibfield  {author} {\bibinfo {author} {\bibfnamefont {J.}~\bibnamefont {{van Tilborg}}},\ }\bibfield  {title} {\enquote {\bibinfo {title} {Active plasma lensing for relativistic laser-plasma-accelerated electron beams},}\ }\href {\doibase 10.1103/PhysRevLett.115.184802} {\bibfield  {journal} {\bibinfo  {journal} {Phys. Rev. Lett.}\ }\textbf {\bibinfo {volume} {115}},\ \bibinfo {pages} {184802} (\bibinfo {year} {2015})}\BibitemShut {NoStop}%
\bibitem [{\citenamefont {Stackhouse}(2022)}]{Stackhouse2022}%
  \BibitemOpen
  \bibfield  {author} {\bibinfo {author} {\bibfnamefont {J.}~\bibnamefont {Stackhouse}},\ }\bibfield  {title} {\enquote {\bibinfo {title} {Preparations on the {BELLA PW} second beamline for staging experiments},}\ }in\ \href {\doibase 10.1109/AAC55212.2022.10822940} {\emph {\bibinfo {booktitle} {2022 IEEE Advanced Accelerator Concepts Workshop (AAC)}}}\ (\bibinfo  {publisher} {IEEE},\ \bibinfo {year} {2022})\BibitemShut {NoStop}%
\bibitem [{\citenamefont {Zingale}(2021{\natexlab{a}})}]{Zingale2021}%
  \BibitemOpen
  \bibfield  {author} {\bibinfo {author} {\bibfnamefont {A.}~\bibnamefont {Zingale}},\ }\bibfield  {title} {\enquote {\bibinfo {title} {Emittance preserving thin film plasma mirrors for {GeV} scale laser plasma accelerators},}\ }\href {\doibase 10.1103/PhysRevAccelBeams.24.121301} {\bibfield  {journal} {\bibinfo  {journal} {Physical Review Accelerators and Beams}\ }\textbf {\bibinfo {volume} {24}},\ \bibinfo {pages} {121301} (\bibinfo {year} {2021}{\natexlab{a}})}\BibitemShut {NoStop}%
\bibitem [{\citenamefont {Poole}(2015)}]{Poole2015Thesis}%
  \BibitemOpen
  \bibfield  {author} {\bibinfo {author} {\bibfnamefont {P.~L.}\ \bibnamefont {Poole}},\ }\emph {\bibinfo {title} {Liquid Crystals as High Repetition Rate Targets for Ultra Intense Laser Systems}},\ \href@noop {} {\bibinfo {type} {{Ph.D.} thesis}},\ \bibinfo  {school} {The Ohio State University} (\bibinfo {year} {2015})\BibitemShut {NoStop}%
\bibitem [{\citenamefont {Poole}(2016{\natexlab{a}})}]{Poole2016APL}%
  \BibitemOpen
  \bibfield  {author} {\bibinfo {author} {\bibfnamefont {P.~L.}\ \bibnamefont {Poole}},\ }\bibfield  {title} {\enquote {\bibinfo {title} {Moderate repetition rate ultra-intense laser targets and optics using variable thickness liquid crystal films},}\ }\href {\doibase 10.1063/1.4964841} {\bibfield  {journal} {\bibinfo  {journal} {Applied Physics Letters}\ }\textbf {\bibinfo {volume} {109}},\ \bibinfo {pages} {151109} (\bibinfo {year} {2016}{\natexlab{a}})}\BibitemShut {NoStop}%
\bibitem [{\citenamefont {Poole}(2016{\natexlab{b}})}]{Poole2016}%
  \BibitemOpen
  \bibfield  {author} {\bibinfo {author} {\bibfnamefont {P.~L.}\ \bibnamefont {Poole}},\ }\bibfield  {title} {\enquote {\bibinfo {title} {Experiment and simulation of novel liquid crystal plasma mirrors for high contrast, intense laser pulses},}\ }\href {\doibase 10.1038/srep32041} {\bibfield  {journal} {\bibinfo  {journal} {Scientific Reports}\ }\textbf {\bibinfo {volume} {6}},\ \bibinfo {pages} {32041} (\bibinfo {year} {2016}{\natexlab{b}})}\BibitemShut {NoStop}%
\bibitem [{\citenamefont {Czapla}(2022)}]{Czapla2022}%
  \BibitemOpen
  \bibfield  {author} {\bibinfo {author} {\bibfnamefont {N.}~\bibnamefont {Czapla}},\ }\emph {\bibinfo {title} {Intense Laser-Plasma Interactions in Ultrathin Films: Plasma Mirrors, Relativistic Effects, and Orbital Angular Momentum}},\ \href {http://rave.ohiolink.edu/etdc/view?acc_num=osu1658486928321502} {Ph.D. thesis},\ \bibinfo  {school} {The Ohio State University} (\bibinfo {year} {2022}),\ \bibinfo {note} {ohioLINK Electronic Theses and Dissertations Center}\BibitemShut {NoStop}%
\bibitem [{\citenamefont {Czapla}(2025)}]{Czapla2025}%
  \BibitemOpen
  \bibfield  {author} {\bibinfo {author} {\bibfnamefont {N.}~\bibnamefont {Czapla}},\ }\bibfield  {title} {\enquote {\bibinfo {title} {A renewable double plasma mirror for {Petawatt}-class lasers},}\ }\href {\doibase 10.1038/s41598-025-07016-3} {\bibfield  {journal} {\bibinfo  {journal} {Scientific Reports}\ }\textbf {\bibinfo {volume} {15}},\ \bibinfo {pages} {21115} (\bibinfo {year} {2025})}\BibitemShut {NoStop}%
\bibitem [{\citenamefont {Schumacher}(2017)}]{Schumacher2017}%
  \BibitemOpen
  \bibfield  {author} {\bibinfo {author} {\bibfnamefont {D.~W.}\ \bibnamefont {Schumacher}},\ }\bibfield  {title} {\enquote {\bibinfo {title} {Liquid crystal targets and plasma mirrors for laser based ion acceleration},}\ }\href {\doibase 10.1088/1748-0221/12/04/C04023} {\bibfield  {journal} {\bibinfo  {journal} {Journal of Instrumentation}\ }\textbf {\bibinfo {volume} {12}},\ \bibinfo {pages} {C04023} (\bibinfo {year} {2017})}\BibitemShut {NoStop}%
\bibitem [{\citenamefont {Zingale}(2021{\natexlab{b}})}]{ZingalePhD}%
  \BibitemOpen
  \bibfield  {author} {\bibinfo {author} {\bibfnamefont {A.}~\bibnamefont {Zingale}},\ }\emph {\bibinfo {title} {Optical Response of Plasmas from Moderate Intensity to the Relativistic Regime}},\ \href {http://rave.ohiolink.edu/etdc/view?acc_num=osu1626345592829131} {Ph.D. thesis},\ \bibinfo  {school} {The Ohio State University} (\bibinfo {year} {2021}{\natexlab{b}}),\ \bibinfo {note} {ohioLINK Electronic Theses and Dissertations Center}\BibitemShut {NoStop}%
\bibitem [{\citenamefont {Chaban}(2020)}]{Chaban}%
  \BibitemOpen
  \bibfield  {author} {\bibinfo {author} {\bibfnamefont {I.}~\bibnamefont {Chaban}},\ }\bibfield  {title} {\enquote {\bibinfo {title} {Crystalline-like ordering of 8cb liquid crystals revealed by time-domain brillouin scattering},}\ }\href {\doibase 10.1063/1.5135982} {\bibfield  {journal} {\bibinfo  {journal} {The Journal of Chemical Physics}\ }\textbf {\bibinfo {volume} {152}},\ \bibinfo {pages} {014202} (\bibinfo {year} {2020})}\BibitemShut {NoStop}%
\bibitem [{\citenamefont {Singh}(2000)}]{Singh2000PhaseTransitionsInLiquidCrystals}%
  \BibitemOpen
  \bibfield  {author} {\bibinfo {author} {\bibfnamefont {S.}~\bibnamefont {Singh}},\ }\bibfield  {title} {\enquote {\bibinfo {title} {Phase transitions in liquid crystals},}\ }\href@noop {} {\bibfield  {journal} {\bibinfo  {journal} {Physics Reports}\ }\textbf {\bibinfo {volume} {324}},\ \bibinfo {pages} {107} (\bibinfo {year} {2000})}\BibitemShut {NoStop}%
\bibitem [{\citenamefont {Singh}\ and\ \citenamefont {Dunmur}(2002)}]{SinghDunmur2002LiquidCrystalsFundamentals}%
  \BibitemOpen
  \bibfield  {author} {\bibinfo {author} {\bibfnamefont {S.}~\bibnamefont {Singh}}\ and\ \bibinfo {author} {\bibfnamefont {D.~A.}\ \bibnamefont {Dunmur}},\ }\href@noop {} {\emph {\bibinfo {title} {Liquid Crystals: Fundamentals}}}\ (\bibinfo  {publisher} {World Scientific Co Pte Ltd},\ \bibinfo {address} {New Jersey; London; Singapore; Hong Kong},\ \bibinfo {year} {2002})\BibitemShut {NoStop}%
\bibitem [{\citenamefont {Matsuhashi}(2007)}]{Matsuhashi2007}%
  \BibitemOpen
  \bibfield  {author} {\bibinfo {author} {\bibfnamefont {N.}~\bibnamefont {Matsuhashi}},\ }\bibfield  {title} {\enquote {\bibinfo {title} {Structure analysis of 4-octyl-4’-cyanobiphenyl liquid-crystalline free-standing film by molecular dynamics simulation},}\ }\href {\doibase 10.2240/azojomo0234} {\bibfield  {journal} {\bibinfo  {journal} {AZojomo}\ }\textbf {\bibinfo {volume} {3}},\ \bibinfo {pages} {1--12} (\bibinfo {year} {2007})}\BibitemShut {NoStop}%
\bibitem [{\citenamefont {{\"O}zgan}(2011)}]{Ozgan2011Thermal}%
  \BibitemOpen
  \bibfield  {author} {\bibinfo {author} {\bibfnamefont {S.}~\bibnamefont {{\"O}zgan}},\ }\bibfield  {title} {\enquote {\bibinfo {title} {Thermal and spectrophotometric analysis of liquid crystal 8cb/8ocb mixtures},}\ }\href {\doibase 10.1007/s13538-011-0034-1} {\bibfield  {journal} {\bibinfo  {journal} {Brazilian Journal of Physics}\ }\textbf {\bibinfo {volume} {41}},\ \bibinfo {pages} {118--122} (\bibinfo {year} {2011})}\BibitemShut {NoStop}%
\bibitem [{\citenamefont {Nguyen}(2014)}]{rupture2014}%
  \BibitemOpen
  \bibfield  {author} {\bibinfo {author} {\bibfnamefont {T.}~\bibnamefont {Nguyen}},\ }\bibfield  {title} {{\selectlanguage {English}\enquote {\bibinfo {title} {Rupture mechanism of liquid crystal thin films realized by large-scale molecular simulations},}\ }}\href {\doibase 10.1039/c3nr05413f} {\bibfield  {journal} {\bibinfo  {journal} {Nanoscale}\ }\textbf {\bibinfo {volume} {6}},\ \bibinfo {pages} {3083--3096} (\bibinfo {year} {2014})}\BibitemShut {NoStop}%
\bibitem [{\citenamefont {Stannarius}(2019)}]{Stannarius2019}%
  \BibitemOpen
  \bibfield  {author} {\bibinfo {author} {\bibfnamefont {R.}~\bibnamefont {Stannarius}},\ }\bibfield  {title} {\enquote {\bibinfo {title} {Freely suspended smectic films with in-plane temperature gradients},}\ }\href {\doibase 10.1088/1367-2630/ab2673} {\bibfield  {journal} {\bibinfo  {journal} {New Journal of Physics}\ }\textbf {\bibinfo {volume} {21}},\ \bibinfo {pages} {063033} (\bibinfo {year} {2019})}\BibitemShut {NoStop}%
\bibitem [{\citenamefont {Abbott}(1997)}]{Abbott1997}%
  \BibitemOpen
  \bibfield  {author} {\bibinfo {author} {\bibfnamefont {N.~L.}\ \bibnamefont {Abbott}},\ }\bibfield  {title} {\enquote {\bibinfo {title} {Surface effects on orientation of liquid crystals},}\ }\href {\doibase 10.1016/S1359-0294(97)80011-2} {\bibfield  {journal} {\bibinfo  {journal} {Current Opinion in Colloid \& Interface Science}\ }\textbf {\bibinfo {volume} {2}},\ \bibinfo {pages} {76--82} (\bibinfo {year} {1997})}\BibitemShut {NoStop}%
\bibitem [{\citenamefont {J\'{e}r\^{o}me}(1991)}]{Jerome1991}%
  \BibitemOpen
  \bibfield  {author} {\bibinfo {author} {\bibfnamefont {B.}~\bibnamefont {J\'{e}r\^{o}me}},\ }\bibfield  {title} {\enquote {\bibinfo {title} {Surface effects and anchoring in liquid crystals},}\ }\href {\doibase 10.1088/0034-4885/54/3/002} {\bibfield  {journal} {\bibinfo  {journal} {Reports on Progress in Physics}\ }\textbf {\bibinfo {volume} {54}},\ \bibinfo {pages} {391--451} (\bibinfo {year} {1991})}\BibitemShut {NoStop}%
\bibitem [{\citenamefont {Picksley}(2024)}]{picksley2024}%
  \BibitemOpen
  \bibfield  {author} {\bibinfo {author} {\bibfnamefont {A.}~\bibnamefont {Picksley}},\ }\bibfield  {title} {\enquote {\bibinfo {title} {Matched guiding and controlled injection in dark-current-free, 10-gev-class, channel-guided laser-plasma accelerators},}\ }\href {\doibase 10.1103/PhysRevLett.133.255001} {\bibfield  {journal} {\bibinfo  {journal} {Phys. Rev. Lett.}\ }\textbf {\bibinfo {volume} {133}},\ \bibinfo {pages} {255001} (\bibinfo {year} {2024})}\BibitemShut {NoStop}%
\bibitem [{\citenamefont {Doumy}(2004)}]{Doumy2004PlasmaMirror}%
  \BibitemOpen
  \bibfield  {author} {\bibinfo {author} {\bibfnamefont {G.}~\bibnamefont {Doumy}},\ }\bibfield  {title} {\enquote {\bibinfo {title} {Complete characterization of a plasma mirror for the production of high-contrast ultraintense laser pulses},}\ }\href {\doibase 10.1103/PhysRevE.69.026402} {\bibfield  {journal} {\bibinfo  {journal} {Physical Review E}\ }\textbf {\bibinfo {volume} {69}},\ \bibinfo {pages} {026402} (\bibinfo {year} {2004})}\BibitemShut {NoStop}%
\bibitem [{\citenamefont {Vazquez~et al.}()}]{repo}%
  \BibitemOpen
  \bibfield  {author} {\bibinfo {author} {\bibfnamefont {A.}~\bibnamefont {Vazquez~et al.}},\ }\href {\doibase 10.5281/zenodo.20561401} {\bibfield  {journal} {\bibinfo  {journal} {{Data Sets for "Stability and optical quality of "windmill"-formed 8CB liquid crystal films for replenishable plasma mirrors"}}\ }10.5281/zenodo.20561401}\BibitemShut {NoStop}%
\end{thebibliography}%

\end{document}


\pagestyle{empty}
\begin{table*}[t!]
\centering
\small
\caption{Wavefront statistics (mean $\pm$ standard deviation) for liquid crystal films, organized by temperature range, analysis pupil size, and linear speed. 120 film formation attempts were performed per speed; the number of successful data points decreased as speed increased.}
\alttext{Alt Text: Wavefront RMS error and peak-to-valley increase with analysis pupil diameter and wiping speed, and vary least between 21 and 22 degrees Celsius. Total angular fluctuation ranges from approximately 0.3 to 0.7 milliradians.}

\label{tab:SI_wavefront_stats}
\vspace{0.1em}
\resizebox{\textwidth}{!}{%
\begin{tabular}{c|c|c|c|c|c|c|c}
\toprule
\textbf{Temperature Range} & \textbf{Pupil Size} & \textbf{Linear Speed (mm/s)} & \textbf{Peak-to-Valley (nm)} & \textbf{Root-Mean-Squared (nm)} & \textbf{Tilt $x$ ($\mu$rad)} & \textbf{Tip $y$ ($\mu$rad)} & \textbf{Total Angular Fluctuation $\sqrt{\sigma_{x}^2+\sigma_{y}^2}$ ($\mu$rad)} \\
\midrule
20-21 \textdegree C & 2 mm & 2.7 & 106 $\pm$ 15 & 15 $\pm$ 2 & -3 $\pm$ 478 & -361 $\pm$ 291 & 559 \\
 &  & 4.1 & 118 $\pm$ 14 & 17 $\pm$ 2 & 261 $\pm$ 514 & -195 $\pm$ 262 & 577 \\
 &  & 5.4 & 133 $\pm$ 14 & 20 $\pm$ 2 & -166 $\pm$ 349 & -109 $\pm$ 203 & 403 \\
 &  & 6.8 & 135 $\pm$ 14 & 19 $\pm$ 2 & -382 $\pm$ 495 & 233 $\pm$ 304 & 581 \\
 &  & 8.1 & 156 $\pm$ 11 & 22 $\pm$ 2 & 228 $\pm$ 486 & 783 $\pm$ 524 & 714 \\
 &  & 9.5 & 189 $\pm$ 16 & 29 $\pm$ 2 & 1887 $\pm$ 189 & 222 $\pm$ 189 & 267 \\
 &  & 10.8 & 221 $\pm$ 20 & 35 $\pm$ 3 & 1919 $\pm$ 236 & 561 $\pm$ 249 & 343 \\
\cmidrule(lr){2-8}
 & 3 mm & 2.7 & 291 $\pm$ 51 & 39 $\pm$ 8 & -5 $\pm$ 479 & -365 $\pm$ 292 & 561 \\
 &  & 4.1 & 355 $\pm$ 55 & 49 $\pm$ 9 & 261 $\pm$ 515 & -199 $\pm$ 263 & 578 \\
 &  & 5.4 & 402 $\pm$ 50 & 59 $\pm$ 8 & -166 $\pm$ 349 & -111 $\pm$ 203 & 404 \\
 &  & 6.8 & 377 $\pm$ 46 & 57 $\pm$ 7 & -381 $\pm$ 496 & 230 $\pm$ 304 & 581 \\
 &  & 8.1 & 471 $\pm$ 37 & 71 $\pm$ 6 & 229 $\pm$ 485 & 779 $\pm$ 524 & 714 \\
 &  & 9.5 & 637 $\pm$ 42 & 98 $\pm$ 8 & 1887 $\pm$ 189 & 219 $\pm$ 189 & 268 \\
 &  & 10.8 & 736 $\pm$ 63 & 116 $\pm$ 11 & 1921 $\pm$ 235 & 559 $\pm$ 251 & 344 \\
\cmidrule(lr){2-8}
 & 4 mm & 2.7 & 725 $\pm$ 131 & 93 $\pm$ 20 & -10 $\pm$ 479 & -370 $\pm$ 291 & 560 \\
 &  & 4.1 & 885 $\pm$ 138 & 117 $\pm$ 22 & 257 $\pm$ 514 & -204 $\pm$ 263 & 578 \\
 &  & 5.4 & 1010 $\pm$ 123 & 141 $\pm$ 19 & -170 $\pm$ 349 & -117 $\pm$ 203 & 404 \\
 &  & 6.8 & 957 $\pm$ 105 & 135 $\pm$ 17 & -385 $\pm$ 495 & 226 $\pm$ 305 & 581 \\
 &  & 8.1 & 1155 $\pm$ 97 & 172 $\pm$ 15 & 224 $\pm$ 486 & 774 $\pm$ 525 & 715 \\
 &  & 9.5 & 1569 $\pm$ 94 & 239 $\pm$ 18 & 1879 $\pm$ 189 & 211 $\pm$ 188 & 267 \\
 &  & 10.8 & 1812 $\pm$ 161 & 282 $\pm$ 27 & 1912 $\pm$ 236 & 549 $\pm$ 251 & 344 \\
\midrule
21-22 \textdegree C & 2 mm & 2.7 & 108 $\pm$ 12 & 12 $\pm$ 1 & -2636 $\pm$ 390 & 1005 $\pm$ 341 & 518 \\
 &  & 4.1 & 129 $\pm$ 16 & 14 $\pm$ 2 & -2548 $\pm$ 309 & 1512 $\pm$ 334 & 455 \\
 &  & 5.4 & 148 $\pm$ 14 & 15 $\pm$ 2 & -2068 $\pm$ 252 & 1978 $\pm$ 294 & 387 \\
 &  & 6.8 & 147 $\pm$ 9 & 14 $\pm$ 1 & -1931 $\pm$ 316 & 2354 $\pm$ 238 & 396 \\
 &  & 8.1 & 145 $\pm$ 12 & 14 $\pm$ 2 & -1874 $\pm$ 250 & 2582 $\pm$ 223 & 335 \\
 &  & 9.5 & 151 $\pm$ 15 & 14 $\pm$ 1 & -1942 $\pm$ 230 & 2898 $\pm$ 250 & 340 \\
 &  & 10.8 & 163 $\pm$ 14 & 16 $\pm$ 2 & -1685 $\pm$ 254 & 3025 $\pm$ 142 & 291 \\
\cmidrule(lr){2-8}
 & 3 mm & 2.7 & 216 $\pm$ 43 & 24 $\pm$ 8 & -2633 $\pm$ 391 & 1001 $\pm$ 341 & 519 \\
 &  & 4.1 & 267 $\pm$ 56 & 35 $\pm$ 10 & -2543 $\pm$ 308 & 1508 $\pm$ 334 & 454 \\
 &  & 5.4 & 294 $\pm$ 41 & 40 $\pm$ 8 & -2065 $\pm$ 249 & 1974 $\pm$ 295 & 386 \\
 &  & 6.8 & 271 $\pm$ 30 & 38 $\pm$ 6 & -1928 $\pm$ 314 & 2354 $\pm$ 242 & 397 \\
 &  & 8.1 & 275 $\pm$ 46 & 39 $\pm$ 8 & -1870 $\pm$ 248 & 2585 $\pm$ 225 & 335 \\
 &  & 9.5 & 285 $\pm$ 33 & 40 $\pm$ 6 & -1937 $\pm$ 229 & 2899 $\pm$ 249 & 338 \\
 &  & 10.8 & 337 $\pm$ 38 & 49 $\pm$ 6 & -1684 $\pm$ 253 & 3025 $\pm$ 143 & 291 \\
\cmidrule(lr){2-8}
 & 4 mm & 2.7 & 476 $\pm$ 134 & 59 $\pm$ 21 & -2633 $\pm$ 390 & 1005 $\pm$ 341 & 518 \\
 &  & 4.1 & 621 $\pm$ 130 & 85 $\pm$ 24 & -2543 $\pm$ 308 & 1512 $\pm$ 333 & 454 \\
 &  & 5.4 & 700 $\pm$ 97 & 100 $\pm$ 18 & -2066 $\pm$ 249 & 1977 $\pm$ 296 & 386 \\
 &  & 6.8 & 667 $\pm$ 65 & 94 $\pm$ 13 & -1929 $\pm$ 314 & 2357 $\pm$ 242 & 396 \\
 &  & 8.1 & 686 $\pm$ 98 & 98 $\pm$ 19 & -1870 $\pm$ 247 & 2588 $\pm$ 225 & 335 \\
 &  & 9.5 & 718 $\pm$ 84 & 102 $\pm$ 14 & -1939 $\pm$ 229 & 2902 $\pm$ 249 & 338 \\
 &  & 10.8 & 844 $\pm$ 95 & 125 $\pm$ 16 & -1686 $\pm$ 253 & 3028 $\pm$ 143 & 290 \\
\midrule
22-23 \textdegree C & 2 mm & 2.7 & 114 $\pm$ 16 & 15 $\pm$ 2 & 3843 $\pm$ 602 & -551 $\pm$ 297 & 671 \\
 &  & 4.1 & 135 $\pm$ 21 & 18 $\pm$ 3 & 4434 $\pm$ 550 & -504 $\pm$ 258 & 608 \\
 &  & 5.4 & 138 $\pm$ 20 & 18 $\pm$ 3 & 4963 $\pm$ 428 & -272 $\pm$ 311 & 529 \\
 &  & 6.8 & 137 $\pm$ 19 & 20 $\pm$ 2 & 5017 $\pm$ 392 & 241 $\pm$ 261 & 471 \\
 &  & 8.1 & 149 $\pm$ 19 & 21 $\pm$ 2 & 5035 $\pm$ 300 & 1010 $\pm$ 279 & 410 \\
 &  & 9.5 & 204 $\pm$ 33 & 28 $\pm$ 7 & 5494 $\pm$ 364 & 1869 $\pm$ 491 & 611 \\
 &  & 10.8 & 272 $\pm$ 23 & 45 $\pm$ 4 & 6082 $\pm$ 236 & 2555 $\pm$ 349 & 421 \\
\cmidrule(lr){2-8}
 & 3 mm & 2.7 & 268 $\pm$ 58 & 36 $\pm$ 9 & 3851 $\pm$ 601 & -548 $\pm$ 297 & 671 \\
 &  & 4.1 & 331 $\pm$ 75 & 47 $\pm$ 12 & 4440 $\pm$ 548 & -499 $\pm$ 257 & 605 \\
 &  & 5.4 & 329 $\pm$ 73 & 47 $\pm$ 12 & 4967 $\pm$ 430 & -266 $\pm$ 310 & 530 \\
 &  & 6.8 & 367 $\pm$ 48 & 56 $\pm$ 9 & 5023 $\pm$ 392 & 246 $\pm$ 261 & 471 \\
 &  & 8.1 & 414 $\pm$ 46 & 65 $\pm$ 9 & 5040 $\pm$ 300 & 1012 $\pm$ 278 & 409 \\
 &  & 9.5 & 577 $\pm$ 154 & 91 $\pm$ 24 & 5497 $\pm$ 366 & 1871 $\pm$ 491 & 612 \\
 &  & 10.8 & 920 $\pm$ 79 & 147 $\pm$ 15 & 6083 $\pm$ 236 & 2555 $\pm$ 348 & 421 \\
\cmidrule(lr){2-8}
 & 4 mm & 2.7 & 697 $\pm$ 170 & 85 $\pm$ 25 & 3852 $\pm$ 600 & -555 $\pm$ 297 & 670 \\
 &  & 4.1 & 874 $\pm$ 185 & 112 $\pm$ 29 & 4441 $\pm$ 547 & -505 $\pm$ 257 & 604 \\
 &  & 5.4 & 860 $\pm$ 169 & 114 $\pm$ 29 & 4966 $\pm$ 430 & -272 $\pm$ 310 & 530 \\
 &  & 6.8 & 934 $\pm$ 114 & 133 $\pm$ 21 & 5021 $\pm$ 392 & 240 $\pm$ 260 & 470 \\
 &  & 8.1 & 1038 $\pm$ 125 & 155 $\pm$ 21 & 5037 $\pm$ 300 & 1005 $\pm$ 278 & 409 \\
 &  & 9.5 & 1482 $\pm$ 389 & 220 $\pm$ 58 & 5493 $\pm$ 366 & 1861 $\pm$ 487 & 609 \\
 &  & 10.8 & 2336 $\pm$ 206 & 355 $\pm$ 36 & 6078 $\pm$ 235 & 2542 $\pm$ 347 & 419 \\
\midrule
23-24 \textdegree C & 2 mm & 2.7 & 83 $\pm$ 9 & 10 $\pm$ 1 & -2057 $\pm$ 381 & 766 $\pm$ 170 & 417 \\
 &  & 4.1 & 96 $\pm$ 14 & 12 $\pm$ 1 & -1476 $\pm$ 316 & 849 $\pm$ 203 & 376 \\
 &  & 5.4 & 93 $\pm$ 11 & 12 $\pm$ 1 & -1573 $\pm$ 323 & 1049 $\pm$ 129 & 348 \\
 &  & 6.8 & 135 $\pm$ 19 & 18 $\pm$ 3 & -521 $\pm$ 312 & 977 $\pm$ 272 & 414 \\
 &  & 8.1 & 153 $\pm$ 19 & 21 $\pm$ 3 & -252 $\pm$ 324 & 1170 $\pm$ 290 & 435 \\
 &  & 9.5 & 149 $\pm$ 16 & 21 $\pm$ 3 & -236 $\pm$ 212 & 1461 $\pm$ 200 & 291 \\
 &  & 10.8 & 154 $\pm$ 20 & 23 $\pm$ 3 & 103 $\pm$ 336 & 1857 $\pm$ 198 & 391 \\
\cmidrule(lr){2-8}
 & 3 mm & 2.7 & 160 $\pm$ 24 & 19 $\pm$ 3 & -2060 $\pm$ 381 & 771 $\pm$ 170 & 418 \\
 &  & 4.1 & 221 $\pm$ 43 & 29 $\pm$ 7 & -1478 $\pm$ 316 & 853 $\pm$ 204 & 376 \\
 &  & 5.4 & 207 $\pm$ 40 & 26 $\pm$ 6 & -1576 $\pm$ 323 & 1054 $\pm$ 130 & 348 \\
 &  & 6.8 & 339 $\pm$ 50 & 51 $\pm$ 9 & -520 $\pm$ 313 & 981 $\pm$ 273 & 415 \\
 &  & 8.1 & 390 $\pm$ 50 & 59 $\pm$ 9 & -247 $\pm$ 324 & 1177 $\pm$ 291 & 436 \\
 &  & 9.5 & 374 $\pm$ 57 & 57 $\pm$ 10 & -233 $\pm$ 211 & 1469 $\pm$ 200 & 291 \\
 &  & 10.8 & 408 $\pm$ 70 & 64 $\pm$ 11 & 102 $\pm$ 336 & 1863 $\pm$ 198 & 390 \\
\cmidrule(lr){2-8}
 & 4 mm & 2.7 & 363 $\pm$ 70 & 46 $\pm$ 8 & -2062 $\pm$ 381 & 770 $\pm$ 170 & 417 \\
 &  & 4.1 & 529 $\pm$ 109 & 70 $\pm$ 17 & -1481 $\pm$ 316 & 852 $\pm$ 204 & 376 \\
 &  & 5.4 & 484 $\pm$ 103 & 63 $\pm$ 16 & -1580 $\pm$ 323 & 1052 $\pm$ 130 & 349 \\
 &  & 6.8 & 821 $\pm$ 126 & 120 $\pm$ 20 & -521 $\pm$ 314 & 980 $\pm$ 273 & 416 \\
 &  & 8.1 & 955 $\pm$ 127 & 142 $\pm$ 21 & -250 $\pm$ 324 & 1175 $\pm$ 291 & 436 \\
 &  & 9.5 & 920 $\pm$ 141 & 136 $\pm$ 24 & -236 $\pm$ 212 & 1467 $\pm$ 200 & 291 \\
 &  & 10.8 & 1031 $\pm$ 174 & 153 $\pm$ 27 & 100 $\pm$ 336 & 1861 $\pm$ 199 & 390 \\
\bottomrule
\end{tabular}%
}
\end{table*}

\bibliography{aipsamp_v2_unbolded}